\documentclass[aps,prd,groupedaddress,showpacs,nofootinbib,amssymb,balancelastpage,preprintnumbers]{revtex4}

\usepackage[dvipdfmx]{graphicx}
\usepackage{graphicx,bm,color}

\usepackage{amsmath}
\usepackage{amssymb}
\usepackage{amsfonts}
\usepackage{cases}
\usepackage{cancel}
\usepackage{hyperref}
\usepackage{ulem}

\usepackage{here}
\usepackage{tikz}
\usetikzlibrary{matrix}

\begin{document}

\title{
New interpretation of chiral phase transition: Violation of trilemma in QCD
}  

\author{Chuan-Xin Cui}\thanks{{\tt cuicx1618@mails.jlu.edu.cn}}
\affiliation{Center for Theoretical Physics and College of Physics, Jilin University, Changchun, 130012,
China}

\author{Jin-Yang Li}\thanks{{\tt lijy1118@mails.jlu.edu.cn}}
\affiliation{Center for Theoretical Physics and College of Physics, Jilin University, Changchun, 130012,
China}

\author{Shinya Matsuzaki}\thanks{{\tt synya@jlu.edu.cn}}
\affiliation{Center for Theoretical Physics and College of Physics, Jilin University, Changchun, 130012,
China}

\author{Mamiya Kawaguchi}\thanks{{\tt mamiya@ucas.ac.cn}} 
      \affiliation{ School of Nuclear Science and Technology, University of Chinese Academy of Sciences, Beijing 100049, China}

\author{Akio Tomiya}\thanks{{\tt akio@yukawa.kyoto-u.ac.jp}}
\affiliation{RIKEN BNL Research center, Brookhaven National Laboratory, Upton, NY, 11973, USA}
\affiliation{Department of Information Technology, International Professional University of Technology in Osaka, 3-3-1 Umeda, Kita-Ku, Osaka, 530-0001, Japan }

\begin{abstract}  
We find that the chiral phase transition (chiral crossover) in QCD at physical point is triggered by big imbalance among three fundamental quantities essential for the QCD vacuum structure: susceptibility functions for the chiral symmetry, axial symmetry, and the topological charge. 
The balance, dobbed the QCD trilemma, is unavoidably violated when one of the magnitudes among them is highly dominated, or suppressed. 
Based on a three-flavor Nambu-Jona-Lasinio model,  
we explicitly evaluate the amount of violation of the QCD trilemma at physical point, and 
show that the violation takes place   
not only at vacuum, but even in a whole temperature regime including the chiral crossover epoch. 
This work confirms and extends the suggestion recently reported from lattice QCD with 
2 flavors on  
dominance of the axial and topological susceptibilities left in the chiral susceptibility at 
high temperatures. 
It turns out that the imbalance is essentially due to the flavor symmetry violation 
of the lightest three flavors, and the flavor breaking   
specifically brings enhancement of the axial anomaly contribution in the chiral order parameter,  
while the the strength of the axial breaking  and the transition rate of the topological 
charge are fairly insensitive to the flavor symmetry. 
The violation of QCD trilemma and its flavor dependence can be tested by lattice 
simulations with 2 + 1 flavors in the future, and would also give a new guiding principle to explore  
the flavor dependence of the chiral phase transition, such as the Columbia plot, 
including possible extension with external fields.

\end{abstract} 
\maketitle

\section{Introduction}

The chiral phase transition is of importance    
to comprehend the QCD vacuum, and is also essential to 
figure out the origin of mass 
in a view of thermal history of the universe. 
Plenty of studies on the chiral phase transition have extensively been worked out so far through the nonperturbative 
analysis in lattice simulations, and also chiral effective models of QCD. 
However, as argued in the literature~\cite{Shuryak:1993ee,Aoki:2021qws}, 
still, it is not well understood whether the chiral symmetry 
breaking is the most dominant source of the origin of mass, even 
in presence of contamination with the $U(1)_A$ anomaly, and by what mechanism it is restored faster than the $U(1)_A$ symmetry at high temperature.

The order parameter of the chiral symmetry is given by the quark condensate, 
which can alternatively be signaled by  
difference of meson correlation functions 
for the chiral partners: the latter is referred to as an indicator of the chiral breaking strength. 
Though being so simple and well-defined, 
the chiral order parameter 
at physical point is actually involved 
due to finite quark masses, which explicitly break the chiral symmetry. 
Indeed, the chiral symmetry is restored at high temperature only in part, refereed to as 
the chiral crossover ~\cite{Aoki:2006we, Bhattacharya:2014ara}. 
Actually it gets more intricate because the chiral order parameter 
(the indicator of the chiral breaking strength)  couples with the indicator of the axial breaking strength 
and topological features of the QCD vacuum via finite quark masses. 
The latter tagging is captured by a robust relation between 
the indicators for the chiral $SU(2)_L \times SU(2)_R$ symmetry 
and $U(1)_A$ axial symmetry, which is constructed from a set of generic anomalous Ward identities for the three-flavor chiral $SU(3)_L \times SU(3)_R$ symmetry~\cite{Nicola:2016jlj,Kawaguchi:2020qvg} (for more details, see also the next section): 
\begin{align} 
\chi_{\eta- \delta}  =   \chi_{\pi- \delta} + \frac{4}{m_l^2} \chi_{\rm top}  
 \,, \label{WI-def}
\end{align}
where $m_l = m_u = m_d$ is the isospin-symmetric mass for the lightest up and down quarks; 
$\chi_{\eta- \delta}  \equiv \chi_\eta - \chi_\delta $ 
and 
$\chi_{\pi-\delta}  \equiv \chi_\pi - \chi_\delta$ 
are differences of meson susceptibilities related to
the partners for the chiral symmetry ($\chi_\eta$ and $\chi_\delta$) --- an indicator for the strength of 
the chiral $SU(2)$ symmetry breaking --- 
and axial symmetry ($\chi_\delta$ and $\chi_\pi$) 

--- an indicator for the strength of the $U(1)_A$ axial breaking; $\chi_{\rm top}$ is 
the topological susceptibility related to the transition rate of the topological charge carried by the 
QCD $\theta$ vaccua.  
By the chiral $SU(2)$ and axial rotations, 
the meson susceptibilities exchange their partners: 
$\chi_{\eta} \leftrightarrow   \chi_{\delta}$ (chiral)
and 
$\chi_{\pi} \leftrightarrow   \chi_{\delta}$ (axial), 
hence 
$\chi_{\eta- \delta} = 0$ and $\chi_{\pi- \delta}=0$  
are signals of  
restorations of the associated symmetries. 
($\chi_{\rm top} < 0$ and other susceptibilities are positive in our sign convention. See also the next section.) 
Thus Eq.(\ref{WI-def}) dictates coherence of 
the chiral $SU(2)$  symmetry breaking and $U(1)_A$ breaking, 
linked with the transition rate 
of the topological charge, where all the breaking is 
controlled by nonzero quark masses. 
This anomalous Ward identity takes the same form 
even in the decoupling limit of strange quark, i.e., 
in the lightest two-flavor limit.

Equation (\ref{WI-def}) plays the essential role to comprehend 
how the effective restoration of the chiral symmetry is correlated with that of the axial symmetry and the temperature dependence of topological susceptibility  
in real-life QCD. 
This gives a new guideline in a sense of exploring the chiral phase transition constrained by Eq.(\ref{WI-def}), and would provide crucial clues 
to answer the questions posed above.

The lattice QCD simulations with 2 + 1 flavors at  
the physical point have revealed a faster 
drop of $\chi_{\eta- \delta}$ , 
than $\chi_{\pi - \delta}$ 
around and above the pseudo-critical temperature of the chiral crossover~\cite{Bhattacharya:2014ara}.  
In the case of 2 flavors at the chiral limit,
the effective restoration of the chiral and axial symmetry 
 has also been discussed through the meson susceptibilities
\cite{Cohen:1996ng,Cohen:1997hz,Aoki:2012yj}

However, those are based on independent measurements 
of two terms, $\chi_{\eta- \delta}$ and $\chi_{\pi- \delta}$, 
with the constraint of Eq.(\ref{WI-def}) disregarded.

Measurements of $\chi_{\rm top}$ on lattice QCD with 2 + 1 flavors  
at around physical point 
and its temperature dependence have been reported~\cite{Petreczky:2016vrs,Bonati:2018blm,Borsanyi:2016ksw},  
in light of detecting the effective restoration of 
the $U(1)_A$ symmetry assuming the much faster restoration of 
the chiral $SU(2)$ symmetry. 
However, those are also individual observations, 
basically separated from measurements of the chiral 
and axial indicators.
Therefore, it is yet uncovered how the temperature dependence of 
$\chi_{\rm top}$ would correlate with the other two, 
with reflecting the constraint of Eq.(\ref{WI-def}).

In a view of the coherence in Eq.(\ref{WI-def}), 
a recent lattice study with two lightest flavors   
has for the first time shown significant contributions from the axial and topological susceptibilities    
($\chi_{\pi - \delta}$ and $\chi_{\rm top}$) 
left in the chiral susceptibility ($\chi_{\eta-\delta}$)
in the chiral crossover domain~\cite{Aoki:2021qws}. 
This would imply that the faster chiral crossover is triggered 
by a sizable cancellation between 
axial and topological susceptibilities, the two terms 
in the right-hand side of Eq.(\ref{WI-def}).

\begin{figure}[t]
  \begin{center}
   \includegraphics[width=8cm]{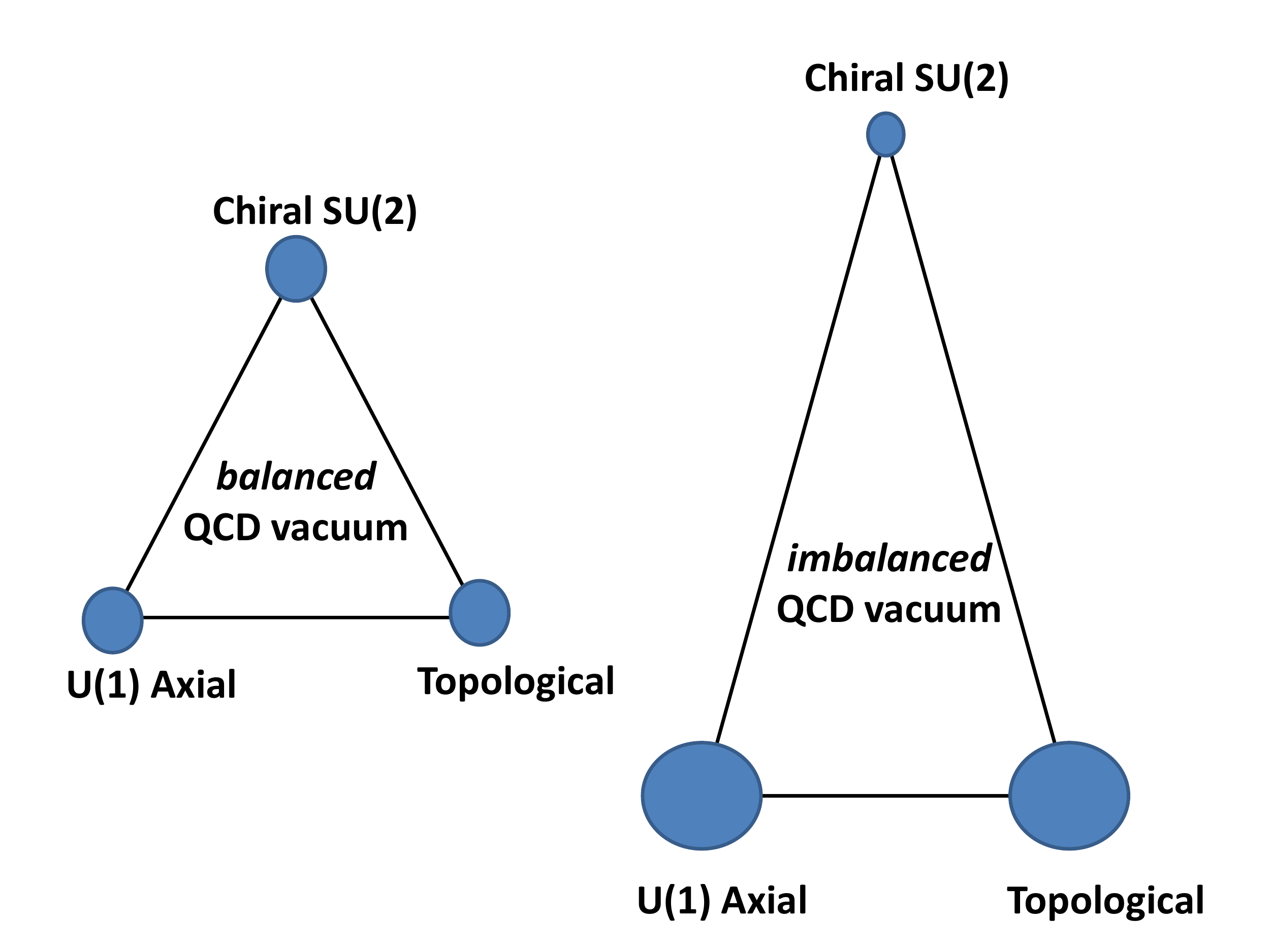}
  \end{center}   
\vspace*{-0.5cm}
\caption{
Illustration of QCD trilemma and its violation.  
The QCD vacuum structure is built upon  
the ``Chiral SU(2)", ``U(1) Axial", and ``Topological" features, 
which are related each other by a balance relation in Eq.(\ref{WI-def})), 
where the ``Chiral SU(2)", ``U(1) Axial" and ``Topological" are 
monitored by $\chi_{\eta - \delta}$, $\chi_{\pi -\delta}$, and $(-4/m_l^2)\cdot \chi_{\rm top}$, 
respectively. 
Left panel: the QCD vacuum is ``balanced" and 
holds the trilemma by forming the equilateral triangle with the same order of the weight amplitudes denoted by blobs. 
Right panel: the trilemma is violated (imbalanced) when 
a big cancellation between 
``U(1) Axial", and ``Topological" takes place 
in Eq.(\ref{WI-def}), which is represented by 
the isosceles triangle with one blob significantly reduced, keeping Eq.(\ref{WI-def}) and the corresponding two sides stretched out. 
As it will turn out in the text, real-life QCD is ``imbalanced". } 
\label{QCD-trilemma}
\end{figure}

To quantify the magnitude of such a cancellation,  
we may define an ideal case with no preference among three 
susceptibilities in magnitude in Eq.(\ref{WI-def}), so that the Ward identity acts like a balance equation. 
We dob this ideal situation as ``QCD trilemma", 
and depict a triangle cartoon in Fig.~\ref{QCD-trilemma}. 
The degree of formation of QCD trilemma can be evaluated via the following 
quantity: 
\begin{align} 
 R &\equiv 
 \frac{\frac{4}{m_l^2} \chi_{\rm top} 
 + \chi_{\pi - \delta}}{\chi_{\eta - \delta} - 
 \frac{4}{m_l^2} \chi_{\rm top}
} 
 = 1 + \frac{\frac{4}{m_l^2} \chi_{\rm top}}{\chi_{\pi - \delta}}
 \,. \label{R-def}
\end{align}
By using this $R$ 
the Ward identity in Eq.(\ref{WI-def}) is rewritten  
as 
\begin{align} 
 \chi_{\eta- \delta} &= R \cdot \chi_{\pi- \delta}  
\,, \notag\\ 
{\rm or} 
\qquad 
 - \frac{4}{m_l^2} \chi_{\rm top} &= (1-R) \cdot  \chi_{\pi - \delta} 
 \,, \label{WI-R}
\end{align}
so that $R$ measures the size of gap in magnitude 
between the chiral ($\chi_{\eta - \delta}$) and axial ($\chi_{\pi- \delta}$) susceptibilities
, or the topological $(\chi_{\rm top})$ and axial ($\chi_{\pi- \delta}$) susceptibilities. 
Equation (\ref{WI-R}) tells us 
that $\frac{\chi_{\eta- \delta}}{\chi_{\pi- \delta}} = R$ and 
$\frac{ \left(- \frac{4}{m_l^2} \chi_{\rm top} \right)}{\chi_{\pi -\delta}} = 1-R$, so  
one may then quantify the amount of balance to keep the QCD trilemma, by 
saying 
that the three susceptibilities are balanced when 
\begin{align} 
{\rm balanced}: \qquad 0.1 < R  < 0.9   
\,, 
\label{natural-region}
\end{align}
otherwise imbalanced along with a big gap by more than one order of magnitude 
between two of three susceptibilities. 
An ideal and optimized trilemma is thus realized when $R=0.5$. 
We shall dub this $R$ as the trilemma estimator. 
$R$ becomes $\ll 1$ when 
axial and topological susceptibilities get close each other with 
different sign in Eq.(\ref{R-def}), which would signal the faster effective restoration of the chiral symmetry than that of 
the axial one through Eq.(\ref{WI-R}).

The aforementioned evidence observed by lattice 
simulations~\cite{Bhattacharya:2014ara} 
on the faster drop of $\chi_{\eta- \delta}$ than $\chi_{\pi - \delta}$ 
indicates $R \ll 1$ in a view of Eq.(\ref{WI-R}). 
The result from the recent lattice study 
with two lightest flavors 
in~\cite{Aoki:2021qws} 
can be rephrased as 
$R \ll 1$ in both Eqs.(\ref{R-def}) and (\ref{WI-R}). 
 Though not explicitly addressed and restricted  
 only around the crossover regime, 
this imbalance could also be read off 
from the existing 
lattice QCD data with 2 + 1 flavors in~\cite{Bhattacharya:2014ara} and also~\cite{Aoki:2021qws} with taking into account possible finite volume effects and statistical errors.

Thus, violation of the QCD trilemma has not been yet explicitly explored at the physical point for   
2 + 1 flavors on the same lattice setting, and it is still unclear how axial and topological susceptibilities, holding single Eq.(\ref{WI-def}) with the chiral one, develop in a whole finite temperature regime and contributes to achieving the chiral crossover. 
Even in the context of effective chiral models, 
no such discussion along with Eq.(\ref{WI-def}) 
has so far been made 
together with proper incorporation of the flavor-singlet condition for $\chi_{\rm top}$~\cite{Baluni:1978rf,Kim:1986ax,Kawaguchi:2020qvg} 
(to the latter point, see also the next section). 
Real-life QCD having 2 + 1 flavors at physical point might be imbalanced in realizing 
the chiral crossover, through undergoing a big cancellation 
between axial and topological susceptibilities 
in a whole temperature regime.

In this paper, 
we discuss the violation of QCD trilemma in real-life QCD   
based on 
a Nambu-Jona-Lasinio (NJL) model, 
and give a qualitative interpretation of the mechanism of 
the violation, namely the coherence among the chiral, axial, 
and topological susceptibilities, constrained by Eq.(\ref{WI-def}). 
Prior to the lattice simulations, 
we show 
that 
real-life QCD  
indeed yields $R \ll 1$, i.e., exhibits the violation of QCD trilemma,    
in a whole temperature regime including the chiral crossover regime. 
We find that 
the violation of QCD trilemma, and 
the related dominance of $\chi_{\pi-\delta}$ and $\chi_{\rm top}$ in the chiral order parameter at the crossover regime 
are due to the three-flavor symmetry violation.

Our findings are shortly testable by lattice simulations, 
and would help deeper understanding of the flavor dependence 
of the chiral phase transition, mapped onto the 
so-called Columbia plot~\cite{Brown:1990ev}. 
Exploring the chiral (crossover) phase transition along 
with the violation of QCD trilemma would lead to 
clues toward answering the posed questions: the expected dominance of the chiral symmetry 
breaking in the origin of mass, and deeper understanding of 
the observed faster (effective) restoration of the chiral symmetry 
in the presence of contamination with the $U(1)_A$ anomaly.

This paper is organized as follows. 
In Sec.~II, we introduce the preliminaries relevant to 
the discussion in the later sections, which include 
definitions and generic formulas for susceptibilities, 
as well as a concise derivation of the anomalous chiral 
Ward identity. 
In Sec.~III, the NJL model that we work throughout this paper 
is introduced, together with showing qualitative consistency of 
the model predictions with the lattice data, which 
includes the temperature dependence of 
the quark condensate, meson susceptibilities, and 
topological susceptibilities. 
In Sec.~IV we discuss the QCD trilemma estimator $R$, 
in a whole temperature region, including the chiral crossover 
regime, and show the violation of the trilemma, imbalance of the 
real-life QCD vacuum. 
We then demonstrate that the violation is due to 
the three-flavor symmetry. 
Sec.~V devotes to our conclusion, where several 
possible applications of the notion of QCD trilemma are also briefly addressed. 

\section{Central formulas: topological susceptibility and anomalous chiral Ward-identities in QCD}

In this section we begin by reviewing the generic expression for the topological susceptibility 
$\chi_{\rm top}$~\cite{Kawaguchi:2020qvg} with 
the flavor singlet condition properly reflected~\cite{Baluni:1978rf,Kim:1986ax}, 
and introduce the related anomalous chiral Ward identities in QCD involving 
pseudoscalar susceptibilities $\chi_{\pi}, \chi_{\eta}$ and $\chi_\delta$.

\subsection{Topological susceptibility: flavor singlet nature}

The topological susceptibility $\chi_{\rm top}$ is related to the $\theta$ vacuum configuration of QCD. It is defined as the curvature of the $\theta$-dependent vacuum energy $V(\theta)$ in QCD at  $\theta=0$: 
	\begin{equation}
		\chi_{\rm top}=-\int_T d^{4}x \frac{\delta^2 V(\theta)}{\delta \theta(x)\delta \theta(0) }\Bigg{|}_{\theta =0}
		\,, \label{chitop:def}
	\end{equation}
where  
the temperature integral $\int_T d^4x $ is defined as $\int_0^{1/T} d\tau \int d^3x $ 
with the imaginary time $\tau = i x_0$, 
and $V(\theta)$ denotes the potential of QCD, which is read off from 
the generating functional of QCD (in Euclidean space):  
\begin{eqnarray}
Z_{\rm QCD}&=&\int [\Pi_f dq_f d\bar q_f] [dA]
\exp\Biggl[-\int_T d^4x \Biggl\{
\sum_{f=u,d,s}\Bigg(
\bar q^f_Li\gamma^\mu D_\mu q^f_L
+
\bar q^f_Ri\gamma^\mu D_\mu q^f_R
\notag\\ 
&& +\bar q^f_L m_f q^f_R+\bar q^f_R m_f q^f_L\Bigg)
+\frac{1}{4g^2}(F_{\mu\nu}^a)^2 
+\frac{i\theta}{32\pi^2}F_{\mu\nu}^a\tilde F_{\mu\nu}^a
\Biggl\}
\Biggl]. 
\label{QCDgene}
\end{eqnarray}
Here 
$q^f_{L(R)}$ denote the left- (right-) handed quark fields; the covariant derivative of the quark field is represented as $D_\mu$ involving the gluon fields $A$; 
$F_{\mu\nu}^a$ is the field strength of the gluon fields with 
$g$ being the QCD coupling constant; for simplicity, 
the quark masses are taken to be real and positive with the electroweak-induced CP violation disregarded.  

The form of the $\theta$ dependence on the QCD generating functional is 
ambiguous, because the $\theta$ parameter can always be shifted by 
the $U(1)_A$ rotation through the induced $U(1)_A$ anomaly as well as the phase shift in the quark mass term. 
\textcolor{black}{Thus the QCD-$\theta$ vacuum is shifted by 
the $U(1)_A$ phase as well. Taking into account this shift, 
the true QCD vacuum is determined so as to set the net $\theta$ to zero, i.e. the CP invariant vacuum, as shown in 
the context of the Vafa-Witten's theorem~\cite{Vafa:1983tf}. 
The topological susceptibility $\chi_{\rm top}$ should then be evaluated at the true vacuum with the net $\theta =0$. This is how to properly compute $\chi_{\rm top}$~\cite{Kim:1986ax}. }

Under the $U_A(1)$ rotation with the rotation angle $\theta_f$, 
the left- and right-handed quark fields are transformed as 
\begin{eqnarray} 
q^f_L&\to& \exp\left( -i\theta_f/2 \right)q^f_L, \nonumber\\
q^f_R&\to& \exp\left( i\theta_f/2 \right)q^f_R\, .  
\end{eqnarray}
We then find that 
the extra phase factor shows up in the QCD generating functional 
written in terms of the transformed chiral quark fields:  
\begin{eqnarray}
&&\int [\Pi_f dq_f d\bar q_f] [dA]
\exp\Biggl[-\int_T d^4x \Biggl\{
\sum_{f=u,d,s}\Bigg(
\bar q^f_Li\gamma^\mu D_\mu q^f_L
+
\bar q^f_Ri\gamma^\mu D_\mu q^f_R
\notag\\ 
&& 
+
\bar q^f_L m_f e^{i\theta_f} q^f_R+\bar q^f_R m_f e^{-i\theta_f} q^f_L\Bigg)
+\frac{1}{4g^2}(F_{\mu\nu}^a)^2
+
\frac{i(\theta-\bar\theta)}{32\pi^2}
F_{\mu\nu}^a\tilde F_{\mu\nu}^a
\Biggl\}
\Biggl],
\end{eqnarray}
where 
$\bar\theta=\sum_{f=u,d,s} \theta_f = \theta_u+\theta_d+\theta_s$.

We can take a phase convention in such a way that 
the $\theta$ dependence goes away from the topological gluonic term $(F \tilde{F})$: 
\begin{eqnarray}
\theta=\bar \theta=\theta_u+\theta_d+\theta_s 
\,. 
\label{choicetheta_f}
\end{eqnarray}
Instead, the quark mass term fully carries the $\theta$-dependence, which makes manifest presence of the strong CP violation. 
Here, the CP violating phases $\theta_{u,d,s}$ actually involves redundancy, i.e., being not fully independent, 
because the QCD interaction and the QCD vacuum characterized by the three-flavor symmetric quark condensate 
are flavor blind (singlet). 
Therefore, we must recover the flavor independence for those CP violating phases. 
Supposing a small enough $\theta_f$ (which is to be consistent with the observation on $\theta < 10^{-10}$ 
from the electric dipole moment of neutron), and expanding the quark mass terms in powers of $\theta_f$s, 
we find that the CP violating coupling terms at the nontrivial leading order arise with the factor of  
$m_f \theta_f$. 
Thus the desired flavor singlet condition goes like~\cite{Baluni:1978rf},
\begin{eqnarray}
m_u\theta_u=m_d\theta_d=m_s\theta_s \equiv x.
\label{singlet_con}
\end{eqnarray}
By using this flavor singlet condition 
together with Eq.~(\ref{choicetheta_f}), $\theta_f$s are determined to be proportional to $\theta$ as
\begin{eqnarray}
\theta_u=\frac{\bar m}{m_u} \theta,\;\;\;
\theta_d=\frac{\bar m}{m_d} \theta,\;\;\;
\theta_s=\frac{\bar m}{m_s} \theta 
\,, 
\label{conspara}
\end{eqnarray}
where
\begin{eqnarray}
\bar m=\left(\frac{1}{m_u}+\frac{1}{m_d}+\frac{1}{m_s} \right)^{-1}.
\end{eqnarray}

Thus the $\theta$-dependent vacuum energy of QCD with the flavor singlet nature properly 
reflected is: 
\begin{eqnarray}
V_{\rm QCD}(\theta)&=&
-\ln\left[
\int [\Pi_f dq_f d\bar q_f] [dA]
\exp\left(-\int_T d^4x 
{\cal L}_{\rm QCD}^{(\theta)}
\right)\right]
\,, \label{V-QCD}
\end{eqnarray}
where
\begin{eqnarray}
{\cal L}_{\rm QCD }^{(\theta)}
=
\sum_f\left(
\bar q^f_Li\gamma^\mu D_\mu q^f_L
+
\bar q^f_Ri\gamma^\mu D_\mu q^f_R
\right)
+
\bar q_L {\cal M}_\theta q_R+\bar q_R {\cal M}_\theta^\dagger q_L
+\frac{1}{4g^2}(F_{\mu\nu}^a)^2
\,,
\label{TQCDlag}
\end{eqnarray}
with 
${\cal M}_\theta$ being the $\theta$-dependent quark matrix,
\begin{eqnarray}
{\cal M}_\theta=
{\rm diag}\left[m_u\exp\left(i\frac{\bar m}{m_u}\theta\right), m_d\exp\left(i\frac{\bar m}{m_d}\theta\right), m_s\exp\left(i\frac{\bar m}{m_s}\theta\right)\right].
\label{calM-theta}
\end{eqnarray} 
Based on Eq.(\ref{V-QCD}), we evaluate $\chi_{\rm top}$ in Eq.(\ref{chitop:def}), and find~\cite{Kawaguchi:2020qvg}
	\begin{align}
		\chi_{\rm top} 
		&=
		 \bar{m}^2 \left[ 
		\frac{\langle \bar {u}u \rangle}{m_l}  
		+\frac{\langle \bar {d}d \rangle}{m_l} 
		+\frac{\langle \bar {s}s \rangle}{m_s} 
		+  
		\chi_{P}^{uu}+\chi_{P}^{dd}
		+\chi_P^{ss} 
		+ 2 \chi_P^{ud} 
		+2\chi_{P}^{us}+
		2\chi_{P}^{ds}
        \right] 
	\notag \\ 
	& = \frac{1}{4} \left[
	m_l\left(\langle \bar{u} u \rangle +\langle \bar{d} d \rangle  \right)
	+ m_l^2\left(\chi_{P}^{uu}+\chi_{P}^{dd}
	+2\chi_{P}^{ud}
	\right)
	\right] 
	= m_s \langle \bar{s}s \rangle + m_s^2 \chi_P^{ss} 
	\,, \label{chitop}
	\end{align}
where the pseudoscalar susceptibilities  
$\chi_P^{uu,dd,ud}$, $\chi_P^{ss}$ and $\chi_P^{us, ds}$ 
are defined as 
\begin{align} 
        \chi_P^{f_1f_2} &=\int_T  d^4 x \langle (\bar q_{f_1}(0) i \gamma _5 q_{f_1}(0))(\bar q_{f_2}(x) i \gamma _5 q_{f_2}(x))\rangle
\,, \qquad {\rm for} \quad {q_{_{f_{1,2}}} = u,d,s} 
\,.  		\label{psesus}
	\end{align} 
In Eq.(\ref{chitop}) we have taken the isospin symmetric limit $m_u=m_d\equiv m_l$.  
The signs of the quark masses and condensates are chosen to be positive and negative, 
respectively, such that $\chi_{\rm top} <0$. 
Note that $\chi_{\rm top} \to 0$, when either of quarks becomes massless ($m_l$ or $m_s$ $\to 0$), reflecting the flavor-singlet nature of the QCD vacuum~\cite{Baluni:1978rf,Kim:1986ax}.

\subsection{Anomalous chiral Ward identities}

The anomalous Ward identities regarding the chiral $SU(3)_L \times SU(3)_R$ symmetry 
are directly read off  
from chiral variations of the QCD potential in Eq.(\ref{V-QCD}). 
The central formula then takes the form 
\begin{align} 
 \langle \delta_a {\cal O}_b(0) \rangle
 = - \int_T d^4 x \langle 
  {\cal O}_b(0) \cdot \bar{q}_f(x) i \gamma_5 \{ T_a, {M} \} q_f(x)   
 \rangle 
 \,, \label{rot-O}
\end{align} 
where  
$T_a = \lambda_a/2$ ($a=1, \cdots, 8$) are generators of $SU(3)$; 
$\delta_a$ stands for the infinitesimal variation of 
the chiral $SU(3)$ transformation associated with the generator $T_a$, under which $q_f$ transforms as 
$\delta_a q_f = i \gamma_5 T_a q_f$; ${\cal O}_b(0)$ 
($b=0, \cdots, 8$) is an arbitrary 
operator. 
In particular, for the pseudoscalar operators 
${\cal O}_b = \bar{q}_f i\gamma_5 T_b q_f$, 
choosing $a=1,2,3, 8$ and $b=0, 8$ with $T_0 = 1/\sqrt{6} \cdot {\bf 1}_{3 \times 3}$, 
we get~\cite{Nicola:2016jlj,Kawaguchi:2020qvg}  
\begin{align} 
\langle \bar uu \rangle +\langle \bar dd \rangle 
 &= - m_l \chi_\pi 
\,, \notag\\  
\langle \bar uu \rangle +\langle \bar dd \rangle
+ 4 \langle \bar s s \rangle
& = 
- \left[ m_l 
\left(\chi_{P}^{uu}+\chi_{P}^{dd}+2\chi_{P}^{ud}\right)
- 2 (m_s + m_l)\left(\chi_{P}^{us}+\chi_{P}^{ds} \right)
+ 4 m_s \chi_P^{ss} \right] 
\,, \notag\\ 
\langle \bar uu \rangle +\langle \bar dd \rangle
-2  \langle \bar s s \rangle
& = 
- \left[ m_l 
\left(\chi_{P}^{uu}+\chi_{P}^{dd}+2\chi_{P}^{ud}\right)
+  (m_l - 2 m_s)\left(\chi_{P}^{us}+\chi_{P}^{ds} \right) 
-  2 m_s \chi_P^{ss} \right] 
		\label{chipi}\,,  
	\end{align} 
where $\chi_{\pi}$ denotes the pion susceptibility defined as 
\begin{align} 
	\chi_{\pi}
	&= \int_T  d^4 x 
	\left[ 
	\langle ( \bar u(0) i \gamma_5  u(0))( \bar u(x) i\gamma_5 u(x))\rangle_{\rm conn}
+ \langle ( \bar d(0)  i\gamma_5 d(0))( \bar d(x) i\gamma_5 d(x))\rangle_{\rm conn}
\right] 
\,,  		\label{chipi-def}
	\end{align} 
with 
$\langle \cdot \cdot \cdot \rangle_{\rm conn}$ being the connected part of the correlation function. 
The form of the anomalous chiral Ward-identities in Eq.(\ref{chipi}) 
will be intact, as long as 
only the quark masses gives the leading order of explicit chiral breaking effects, 
as evident in the chiral variation of Eq.(\ref{rot-O}).

Combining Ward identities in Eq.(\ref{chipi}), we find 
	\begin{equation}
	\begin{aligned}
		\chi_{\rm top} &= \frac{1}{2} m_l m_s
		\left(\chi_{P}^{us}+\chi_{P}^{ds} \right)
		=\frac{1}{4}  m_l^2 ( 
		\chi_\eta 
		-\chi_{\pi})
		\,,  
	\end{aligned}
	\label{disc}
	\end{equation} 
	where $\chi_\eta$ is the eta meson susceptibility, defined as 
\begin{align}
	\chi_{\eta} &=\int_T d^4 x 
	\Bigg[ 
	\langle (\bar u(0) i \gamma _5  u(0))
	(\bar u(x) i\gamma _5 u(x))\rangle 
	+
	\langle (\bar d(0) i\gamma _5  d(0))
	(\bar d(x) i\gamma _5 d(x))\rangle
\notag\\ 
& +
	2 
	\langle (\bar u(0) i\gamma _5  u(0))
	(\bar d(x) i\gamma _5 d(x))\rangle
	\Bigg] 
	\notag\\ 
	& = \chi_{P}^{uu}+\chi_{P}^{dd}+2\chi_{P}^{ud}
	\,. \label{chi-eta}
\end{align} 
The last line of Eq.(\ref{disc}) can be written as  
\begin{align} 
(\chi_\eta - \chi_\delta)  =   (\chi_\pi - \chi_\delta) + \frac{4}{m_l^2} \chi_{\rm top}  
 \,, \label{WI-def-app}
\end{align}
where $\chi_\delta$ is the susceptibility for the delta meson channel ($a_0$ meson in terms of the Particle Data Group  identification), defined in the same way as $\chi_\pi$ in Eq.(\ref{psesus}) 
with the factors of $(i \gamma_5)$ replaced with identity $1$. 
$\chi_{\eta- \delta}  \equiv \chi_\eta - \chi_\delta $ 
and 
$\chi_{\pi-\delta}  \equiv \chi_\pi - \chi_\delta$ 
play the roles of the indicators to detect the strength of the chiral and axial breaking, which signal 
the restorations when those (asymptotically) reach zero.   

Equation (\ref{WI-def-app}) is our central formula, which has been inferred in Eq.(\ref{WI-def}), 
and will be explored in details in the later section.

\section{A chiral effective model: NJL}
In this section we introduce an  NJL model that we work on, 
and give a couple of preliminaries for discussion on the 
estimate of the QCD trilemma estimator $R$ in Eq.(\ref{R-def}), 
with showing consistency of the NJL estimates with currently 
available lattice data on 2 + 1 flavors at physical point. 
Since the methodology to compute observables and thermodynamic quantities 
in the model is standard and fully described in a review~\cite{Hatsuda:1994pi}, 
we will skip all the details, and just present the final formulas directly used 
to the numerical evaluation of the quark condensates and susceptibilities.

The three-flavor NJL model Lagrangian that we work on is constructed as follows:
	\begin{align}
		\mathcal{L}& =\bar{q}(i\gamma_{\mu}\partial^{\mu} -{\bf m} )q+\mathcal{L}_{4f}+\mathcal{L}_{\rm KMT} \,, \notag \\
		\mathcal{L}_{4f}&=\frac{g_s}{2}\sum^{8}_{a=0}\lbrack(\bar{q}\lambda^a q)^2+(\bar{q}i \gamma_5 \lambda^a q )^2\rbrack \,, \notag \\
		\mathcal{L}_{\rm KMT}&=g_D\lbrack \mathop{\rm det}\limits_{i,j}\bar{q_i}(1+\gamma_5 )q_j+ {\rm h.c.} \rbrack 
		\,, \label{Lag:NJL}
	\end{align}
	where $q$ is the $SU(3)$ triplet-quark field, $q=(u,d,s)^T$. 
	The current quark masses are embedded in the mass matrix $\bf m$ of the form ${\bf m}={\rm diag}(m_u, m_d, m_s)$. 

	The four-fermion interaction term  $\mathcal{L}_{4f}$ 
	is invariant under the chiral $U(3)_L\times U(3)_R$  transformation: $q \to U \cdot q$ with $U= \exp[ - i \gamma_5 \sum_{a=0}^8  (\lambda^a/2) \theta^a ]$ and the chiral phases $\theta^a$. 
	The mass term in ${\cal L}$ explicitly breaks $U(3)_L\times U(3)_R$ symmetry. 
	The determinant term $\mathcal{L}_{\rm KMT}$, called  the Kobayashi- Maskawa-‘t Hooft \cite{Kobayashi:1970ji,Kobayashi:1971qz,tHooft:1976rip,tHooft:1976snw} term, induced from 
	the QCD instanton configuration, 
	preserves $SU(3)_L\times SU(3)_R$ invariance (associated with the chiral phases labeled as $a=1, \cdots, 8$) but breaks the $U(1)_A$ (corresponding to $a=0$) symmetry, measured by the effective coupling constant $g_D$.

The $U(1)_A$ symmetry is anomalous due to not only the quark mass terms, but also the KMT term reflecting the underlying gluonic anomaly. Thus we have the anomalous conservation law: 
\begin{eqnarray}
\partial^\mu j_\mu^{a=0} &=&2i \bar q {\bm m}\gamma_5 q
-12 g_D {\rm Im} \left[{\rm det}\bar q_i(1-\gamma_5)q_j  \right]
\, .
\label{current_cons_law_NJL}
\end{eqnarray}
\textcolor{black}{
The matching with the 
underlying QCD leads to an operator relation: 
$Q =  - 4 g_D {\rm Im} \left[{\rm det}\bar q_i(1-\gamma_5)q_j  \right] $, 
where 
$Q=g^2/(32\pi^2)\, F_{\mu\nu}^a \tilde F^{a\mu\nu}$ is the topological charge. 
Using this operator relation together with 
Eq.(\ref{chitop:def}), 
one could evaluate $\chi_{\rm top}$ 
as in the literature, e.g., Refs.~\cite{Fukushima:2001hr,Costa:2008dp,Jiang:2012wm,Jiang:2015xqz} in the framework of the NJL model with the mean field approximation. However, this procedure makes the flavor-singlet nature of the vacuum nontransparent. 
The identification of $\chi_{\rm top}$ defined within the NJL model with the mean field approximation with that derived directly in QCD 
requires a careful separate investigation ensuring the flavor
singlet nature of the vacuum in both theories. 
Instead, in the present paper the NJL-model 
is considered as a reduction of the full QCD allowing the evaluation 
of the right-hand side of Eq.(\ref{chitop})  
in which the flavor singlet nature is manifestly built-in. 
} 


	The NJL model itself is a (perturbatively) nonrenormalizable field theory because ${\cal L}_{4f}$ and ${\cal L}_{\rm KMT}$ describe the higher dimensional 
    interactions with mass dimension greater than four.  
    Therefore, a momentum cutoff $\Lambda$ must be introduced to make the NJL model regularized. 
    We adapt a sharp cutoff regularization for three-dimensional momentum integration, following 
    the literature~\cite{Hatsuda:1994pi}.

\subsection{Gap equations}
%
	 
	 We employ the mean-field approximation, corresponding to the large $N_c$ limit, 
	 and then derive the gap equation and the thermodynamic potential~\cite{Hatsuda:1994pi}. 
	 There the quark condensates (on thermal average) act as the variable of the potential and are $T$-dependent, which we define as 
	 \begin{equation}
	 	\langle \bar u u \rangle \equiv \alpha, 
	 	\quad \langle \bar d d \rangle \equiv \beta, 
	 	\quad \langle \bar s s \rangle \equiv \gamma\,. 
	 \end{equation} 
 Searching for the minimum point of the thermodynamic potential with respect to $\alpha$, $\beta$, and $\gamma$ as variational parameters, we find the stationary conditions, 
 corresponding to the gap equations~\cite{Kunihiro:1983ej,Hatsuda:1994pi}: 
\begin{equation}
		\langle  \bar q_i q_i \rangle =- 2N_c\int^{\Lambda}\frac{d^3 p }{(2\pi)^3}  \frac{M_i}{E_{i}}\big{[}1-2 (\exp (E_{i}/T)+1)^{-1}\big{]} , 
		\label{qcond}
\end{equation}
where $E_{i}=\sqrt{M_i^2+p^2}$, $N_c$ denotes the number of colors to be fixed to three, 
and $M_i$ are full quark masses including the dynamically generated terms:
	\begin{align} 
		M_u&=m_u-2g_s \alpha - 2g_D\beta \gamma \, \notag\\
		M_d&=m_d-2g_s \beta - 2g_D\alpha \gamma \, \notag\\
		M_s&=m_s-2g_s \gamma - 2g_D\alpha \beta \,. 
		\label{fullmasses}
	\end{align}

\subsection{Chiral and axial susceptibilities}

 In this subsection, we introduce susceptibilities for pseudoscalar and scalar meson channels and give their explicit formulas in the present NJL model. 

\subsubsection{Pseudoscalar meson channel}

In the $
\eta$ - $\eta \prime$ coupled channel, the pseudoscalar meson susceptibility is defined on  the generator basis as 
\begin{equation}
	\chi_P^{ij} = \int_T d^4 x \langle (i \bar q (x) \gamma_5 \lambda^i  q (x))(i \bar q (0)\gamma_5 \lambda^j  q(0)) \rangle\,, 
\end{equation}
 where $i,j=0,8$. 
This $\chi_P^{ij}$ takes a matrix form 
\begin{equation}
	\chi_{P}=\frac{-1}{1+G_{P}\Pi_{P}(0,0)} \cdot \Pi_{P}(0,0)
\,, \label{chitop_gen}
\end{equation}
where $G_P$ is the coupling strength matrix and $\Pi_P$ is the polarization tensor matrix, which are given respectively as 
\begin{equation}
G_P=
\begin{pmatrix}
	G_P^{00} & G_P^{08}\\
	G_P^{80} & G_P^{88}
\end{pmatrix}
=
\begin{pmatrix}
	g_s-\frac{2}{3}(\alpha +\beta+\gamma )g_D &  -\frac{\sqrt{2}}{6} (2\gamma-\alpha-\beta )g_D \\
	-\frac{\sqrt{2}}{6} (2\gamma-\alpha-\beta )g_D & g_s-\frac{1}{3}(\gamma-2\alpha -2\beta )g_D
\end{pmatrix}
\,, \label{Gs-p}
\end{equation}
\begin{equation}
\Pi_P=
\begin{pmatrix}
	\Pi _P^{00} & \Pi _P^{08}\\
	\Pi _P^{80} & \Pi _P^{88}
\end{pmatrix}
=
\begin{pmatrix}
	\frac{2}{3}	(2I_P^{uu}+I_P^{ss}) &  \frac{2\sqrt{2}}{3}(I_P^{uu}-I_P^{ss}) \\\frac{2\sqrt{2}}{3}(I_P^{uu}-I_P^{ss} ) & \frac{2}{3} (I_P^{uu} + 2I_P^{ss})
\end{pmatrix}
\,, \label{Pi-p}
\end{equation}
with $I_P^{ii}(\omega, \boldsymbol{p})$ being the pesudoscalar one-loop polarization functions~\cite{Kunihiro:1991hp}, 
\begin{equation}
	I_P^{ii}(0,0)=-\frac{N_c}{\pi ^2}\int^\Lambda _0  d p\, p^2  \frac{1}{E_{i}}\left[1 -2 \left( \exp(E_i /T)+1\right)^{-1} \right]\,, \qquad {\rm for } \quad i=u,d,s 
	\,. \label{IPij}
\end{equation}

 By performing the basis transformation, 
the pseudoscalar susceptibilities defined in Eq.(\ref{psesus}) on the flavor basis are thus obtained as 
\begin{equation}
	\begin{pmatrix}
	\frac{1}{2}\chi_P^{uu}
	+\frac{1}{2}\chi_P^{ud} = \frac{1}{4} \chi_\eta
 	\\
	\chi_P^{us} \\
	\chi_P^{ss}
	\end{pmatrix}
=
	\begin{pmatrix}
	\frac{1}{6} & \frac{\sqrt2}{6} & \frac{1}{12} \\
    \frac{1}{6} & -\frac{\sqrt2}{12} & -\frac{1}{6} \\
    \frac{1}{6} & -\frac{\sqrt2}{3} & \frac{1}{3}
	\end{pmatrix}
	\begin{pmatrix}
	\chi_P^{00} \\
	\chi_P^{08} \\
	\chi_P^{88}
	\end{pmatrix}
	\label{chip}
	\,, 
\end{equation}
where we have taken the isospin symmetric limit into account, i.e., $\chi_P^{uu} = \chi_P^{dd}$ and $\chi_P^{us}=\chi_P^{ds}$.

For $\chi_\pi$ defined in Eq.(\ref{psesus}), 
the explicit formula in the NJL model reads~\cite{Hatsuda:1994pi} 
\begin{equation}
	\chi_{\pi}=\frac{-1}{1+G_{\pi}\Pi_{\pi}(0,0)} \cdot \Pi_{\pi}(0,0)
	\label{pion}
\,, 
\end{equation}
where $G_{\pi} = g_s + g_D \gamma$, which is the coupling strength in the pion channel, and $\Pi_{\pi}$ is the quark-loop polarization function for $\chi_{\pi}$, which is evaluated 
by using $I_P^{ii}$ in Eq.(\ref{IPij}) as  
\begin{equation}
	\Pi_{\pi}=I_P^{uu}+I_P^{dd}=2I_P^{uu}
\,. 
\end{equation}

\subsubsection{Scalar meson channel}

The definitions of scalar susceptibilities are similar to those for pseudoscalars', which are given just by removing {\it{$i\gamma_5$}} in the definition of pseudoscalar susceptibilities, 
and supplying  
the appropriate one-loop polarization functions and the corresponding coupling constants.

In the $0$ - $8$ coupled channel, the scalar susceptibility matrix $\chi_S$ is evaluated in the 
present NJL on the generator basis as 
\begin{equation}
	\chi_{S}=\frac{-1}{1+G_{S}\Pi_{S}(0,0)} \cdot \Pi_{S}(0,0)
\,, \label{chi-S}
\end{equation}
where $G_{S}$ is the coupling strength matrix, 
\begin{equation}
	G_S=
\begin{pmatrix}
	G_S^{00} & G_S^{08}\\
	G_S^{80} & G_S^{88}
\end{pmatrix}
=
\begin{pmatrix}
	g_s+\frac{2}{3}(\alpha +\beta+\gamma )g_D &  \frac{\sqrt{2}}{6} (2\gamma-\alpha-\beta )g_D \\
	\frac{\sqrt{2}}{6} (2\gamma-\alpha-\beta )g_D & g_s+\frac{1}{3}(\gamma-2\alpha -2\beta )g_D
\end{pmatrix}
\,. \label{Gs-s}
\end{equation}
The scalar polarization tensor matrix $\Pi_S$ in Eq.(\ref{chi-S}) is given by  
\begin{equation}
\Pi_S=
\begin{pmatrix}
	\Pi _S^{00} & \Pi _S^{08}\\
	\Pi _S^{80} & \Pi _S^{88}
\end{pmatrix}
=
\begin{pmatrix}
	\frac{2}{3}	(2I_S^{uu}+I_S^{ss}) &  \frac{2\sqrt{2}}{3}(I_S^{uu}-I_S^{ss}) \\\frac{2\sqrt{2}}{3}(I_S^{uu}-I_S^{ss} ) & \frac{2}{3} (I_S^{uu} + 2I_S^{ss})
\,, 
\end{pmatrix}
\end{equation}
with the integral functions,  
\begin{equation}
	I_S^{ii}(0,0)=-\frac{N_c}{\pi ^2}\int^\Lambda _0 p^2 dp \frac{E_{i}^2-M_i^2}{E_i^3}\lbrace1-2[\exp(E_i /T)+1]^{-1}\rbrace \qquad i=u,d,s
\,. 
\end{equation}
By moving on to the flavor base via the base transformation, 
the scalar susceptibilities are cast into the form:
\begin{equation}
	\begin{pmatrix}
	\frac{1}{2}\chi_S^{uu}
	+\frac{1}{2}\chi_S^{ud}
	 \\
	\chi_S^{us} \\
	\chi_S^{ss}
	\end{pmatrix}
=
	\begin{pmatrix}
	\frac{1}{6} & \frac{\sqrt2}{6} & \frac{1}{12} \\
    \frac{1}{6} & -\frac{\sqrt2}{12} & -\frac{1}{6} \\
    \frac{1}{6} & -\frac{\sqrt2}{3} & \frac{1}{3}
	\end{pmatrix}
	\begin{pmatrix}
	\chi_S^{00} \\
	\chi_S^{08} \\
	\chi_S^{88}
	\end{pmatrix}
\,, 
\end{equation}
in which we have read $\chi_S^{uu}=\chi_S^{dd}$
and $\chi_S^{us}=\chi_S^{ds}$. 
From this, the $\sigma$ meson susceptibility can be also read off as 
\begin{align}
	\chi_{\sigma} &=\int_T d^4 x 
	\left[ 
	\langle (\bar u(0)  u(0))
	(\bar u(x) u(x))\rangle 
	+
	\langle (\bar d(0)   d(0))
	(\bar d(x)  d(x))\rangle
	+
	2 
	\langle (\bar u(0)  u(0))
	(\bar d(x)  d(x))\rangle
	\right] \notag \\
&= 2 \chi_S^{uu}+2 \chi_S^{ud}
	\label{sigma}
\,. 
\end{align}
We will not directly evaluate this $\chi_{\sigma}$ in the later section, 
but it will be inferred when the ambiguity in subtracting 
the original form of the Ward-identity in Eq.(\ref{disc}) by 
the scalar meson susceptibility, to get the relation 
between the chiral and axial indicators, and the topological susceptibility 
(in the later section, Sec.~4).

For the $\delta$ meson susceptibility, it is defined as 
\begin{equation}
	\chi_{\delta} 
	=\int_T d^4 x 
	\left[ 
	\langle ( \bar u(0)  u(0))( \bar u(x) u(x))\rangle_{\rm conn}
+ \langle ( \bar d(0)  d(0))( \bar d(x) d(x))\rangle_{\rm conn} 
\right]
\,. 
\end{equation}
 Similar to $\chi_\pi$ in Eq.(\ref{pion}), 
 the explicit formula for $\chi_\delta$ reads~\cite{Hatsuda:1994pi} 
\begin{equation} 
	\chi_{\delta}=\frac{-\Pi_{\delta}(0,0)}{1+G_{\delta}\Pi_{\delta}(0,0)}
	\label{delta}
	\,, 
\end{equation}
where $G_{\delta} = g_s - g_D \gamma$, which is the coupling strength in the $\delta$ channel, and $\Pi_{\delta}=I_S^{uu}+I_S^{dd}=2I_S^{uu}$ is the corresponding quark-loop polarization function. 

\subsection{Topological susceptibility in NJL model}

Combining Eq.~\ref{disc}, \ref{chitop_gen}, \ref{Gs-p}, and \ref{chip}, we get the formula of $\chi_{\rm top}$ written in 
terms of the present NJL-model parameters: 
\begin{align}
	\chi_{\rm top}&=\frac{1}{2}m_l m_s (\chi_p^{us}+\chi_p^{ds}) \notag \\ 
	&=m_l m_s\left( \frac{1}{6}\chi_P^{00} - \frac{\sqrt{2}}{12}\chi_{P}^{08} -\frac{1}{6}\chi_{P}^{88} \right) \notag \\
	&=\frac{-m_l m_s}{\rm{det}(1+G_P \Pi_P)} \bigg{[} \frac{1}{6}(\Pi^{00}_P \Pi^{88}_P G^{88}_P -{\Pi^{08}_P}^2 G^{88}_P ) - \frac{\sqrt{2}}{12} ({\Pi^{08}_P}^2 G^{08}_P 
	-\Pi^{00}_P \Pi^{88}_P G^{08}_P) -\frac{1}{6}(\Pi^{88}_P \Pi^{00}_P G^{00}_P - {\Pi^{80}_P}^2 G^{00}_P)  \bigg{]} 
	\notag \\
	&= \frac{-m_l m_s g_D \alpha}{6 \, \rm{det}(1+G_P \Pi_P)} ((\Pi^{80}_P)^2-\Pi^{88}_P \Pi^{00}_P)
\,. 
\end{align}
One can clearly see that $\chi_{\rm top} \propto m_l m_s$, which is reflected by the flavor-singlet nature of the QCD vacuum. Note also that $\chi_{\rm top} \propto g_D$. 
This should be so because the KMT determinant coupling $g_D$ is directly linked with the gluonic anomaly term in the axial anomaly equation (\ref{current_cons_law_NJL}). 

\subsection{NJL estimates}

In this subsection, we evaluate the temperature dependence of the (subtracted) quark condensate, scalar and pseudoscalar susceptibilities, and the topological susceptibility. 
We also check the consistency with the recent lattice QCD data on 2 + 1 flavors at physical point, and also with other effective models of QCD.

\subsubsection{Parameter setting}

In the present NJL model of Eq.(\ref{Lag:NJL}), we have five parameters that need to be fixed: the light quark mass $m_l$, the strange quark mass $m_s$, the coupling constants $g_s$ and $g_D$, and the three-momentum cutoff $\Lambda$. 
To fix the parameters, we take the following conservative and empirical input values at 
$T=0$ in the isospin symmetric limit~\cite{Hatsuda:1994pi}:
\begin{equation}
	m_{\pi}=136 \, {\rm MeV}, \quad f_{\pi}=93 \, {\rm MeV}, 
	\quad m_K=495.7 \, {\rm MeV}, \quad m_{\eta \prime}=957.5\, {\rm MeV} 
\,. \label{inputs}
\end{equation}
For readers convenience, 
concise NJL formulae are presented in Appendix~\ref{App:formulas}. 
To fix the remaining one degree of freedom, we follow the literature~\cite{Miller:1990iz,Hatsuda:1994pi} to take light quark mass $m_l=5.5$ MeV (at the renormalization scale of 1 GeV). Thus all the model  parameters are fixed, which are presented in Table~\ref{Parameter}. 

\begin{table}[t]
\centering
 
\begin{tabular}{|l|c|}
\hline
model parameter description & input value               \\ 
\hline \hline 
isospin symmetric light quark mass $m_l$   & 5.5 MeV                       \\ 
\hline
strange quark mass $m_s$  & 138 MeV                       \\ \hline
four-fermion coupling constant $g_s$                & 0.358 ${\rm fm}^2$   \\ 
\hline
six-fermion coupling constant $g_D$                & $-$ 0.0275 ${\rm fm}^5$ \\ 
\hline
cutoff $\Lambda$             & 631.4 MeV                     \\ \hline
\end{tabular}
\caption{The model parameter setting, followed from  \cite{Hatsuda:1994pi}.}
\label{Parameter}
\end{table}

With the above parameter set~\footnote{
The constituent quark masses are also 
estimated as 
$M_u(T=0)=M_d(T=0) \simeq 334.2 {\rm MeV}$, $M_s(T=0) \simeq 530.1 {\rm MeV}$, {\textcolor{black}{where use has been made of 
Eqs.(\ref{qcond}), (\ref{fullmasses}), together with the model parameters 
listed in Table~\ref{Parameter}}. } 
}, we estimate the topological susceptibility 
\begin{align} 
\chi_{\rm top} \simeq 0.025/{\rm fm}^4 
\,. 
\end{align} 
For this $\chi_{\rm top}$, 
comparison with the results from the lattice QCD simulations is 
available, which are  
$\chi_{\rm top}=0.019(9)/{\rm fm}^4$~\cite{Bonati:2015vqz}, and $\chi_{\rm top}=0.0245(24)_{\rm stat}(03)_{\rm flow}(12)_{\rm cont}/{\rm fm}^4$~\cite{Borsanyi:2016ksw}. 
Here, for the latter the first error is statistical, the second one comes from the systematic error, and the third one arises due to changing the upper limit of the lattice spacing range in the fit. 
Although their central values do not agree each other, 
we may conservatively say that the 
difference between them is interpreted as a systematic error from the individual lattice QCD calculation.

We will not consider intrinsic-temperature dependent couplings, 
instead, all the $T$ dependence should be induced only 
from the thermal quark loop corrections to the couplings 
defined and introduced at vacuum. 
As it will turn out below, the present NJL shows quite good 
agreement with lattice QCD results on the temperature scaling 
(normalized to the pseudo-critical temperature) 
for the chiral, axial, and topological susceptibilities. 
In this sense, we do not need to introduce such an 
intrinsic $T$ dependence for the model parameters 
in the regime up to temperatures around the 
chiral crossover.

\subsubsection{Subtracted quark condensate}

The quark condensate in the NJL model involves a ultraviolet divergence (which is dominated by a quadratic divergence) due to its vacuum part ($\langle -\bar{q}q \rangle \sim N_c m_q \Lambda^2/(4 \pi^2)$), 
and is needed to be renormalized when compared with lattice data. Since the quadratic divergences in the quark condensate come along with current quark masses (as above), we use a subtracted quark condensate as the chiral order parameter, which has been adopted in the lattice simulations: 
$\Delta_{l,s}(T)\equiv \langle \bar l l \rangle - \frac{m_l}{m_s}\langle \bar s s \rangle$ , where $\langle \bar l l \rangle =\langle \bar u u \rangle=\langle \bar d d \rangle$.

\begin{figure}[t]
\begin{center} 
    \includegraphics[width=0.6\linewidth]{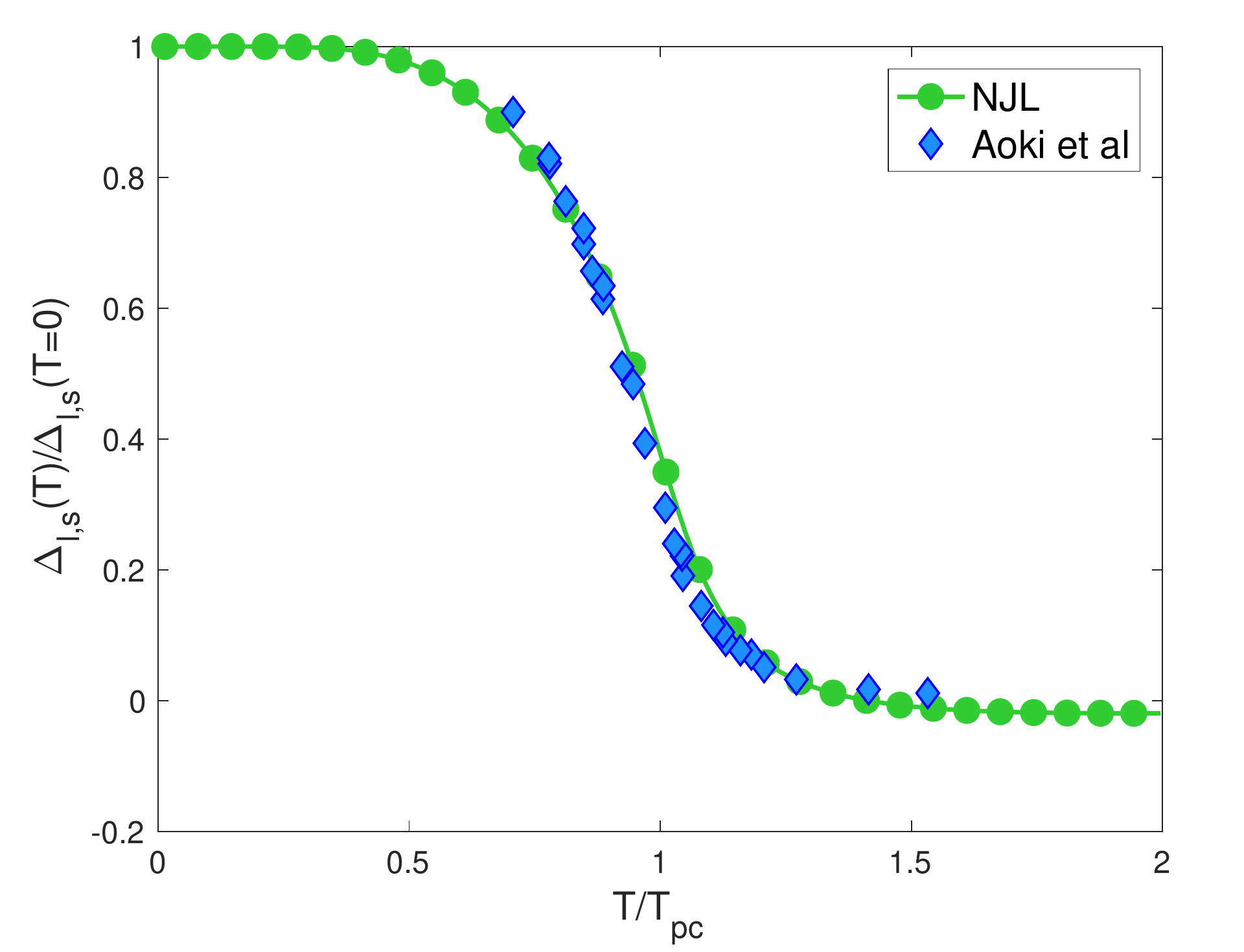}
    \end{center} 
	\caption{ $T/T_{\rm pc}$ dependence of the subtracted quark condensate, in comparison with data from the lattice QCD with 2 + 1 flavors~\cite{Aoki:2009sc}. The normalization factor, 
	the pseudo-critical temperature for the chiral crossover ($T_{\rm pc}$) has been set to individual values estimated from the present NJL model 
($T_{\rm pc}|_{\rm NJL} \simeq 188$ MeV) and the lattice simulation 
($T_{\rm pc}|_{\rm lat.} \simeq 155$ MeV). 
	 }
	\label{subquacond}
\end{figure}

Figure~\ref{subquacond} shows the subtracted quark condensate as a function of temperature predicted from the present NJL model, in comparison with the 2+1 flavor data from the lattice QCD at the physical point~\cite{Aoki:2009sc}. 
The pseudo-critical temperature $T_{\rm pc}$ is (for the NJL prediction) defined 
as $d^2\langle \bar l l \rangle (T) /dT^2|_{T=T_{\rm pc}}=0$. 
We have found $T_{\rm pc}|_{\rm NJL}$  $\simeq 188$ MeV, which is compared with  
the lattice result $T_{\rm pc}|_{\rm lat.}$  $\simeq 155$ MeV~\cite{Aoki:2009sc, Borsanyi:2011bn, Ding:2015ona, Bazavov:2018mes, Ding:2020rtq}.

In the figure, we have normalized $T$ by their $T_{\rm pc}$. 
\textcolor{black}{The reason to take the rescaled dimensionless ratios is to  
make possible systematic errors for the model selection and calculation reduced. 
For instance, the NJL model predicts somewhat larger $T_{pc}$, and $\langle \bar{q}q  \rangle$ at any $T$, 
and basically all dimensionful quantities tend to be overestimated by about 30\%: 
that is thought of as a systematic trend which could be 
associated with the validity of the large $N_c$ approximation. 
Then, the dimensionless quantities like $T/T_{pc}$ and 
$\langle \bar{q}q  \rangle_T/\langle \bar{q}q \rangle_{T=0}$ can have reduced systematic errors. 
So, it would be better to take the rescaled ratios, 
in order to comprehend how the current effective model can be 
compatible with the lattice result. }

From Fig.~\ref{subquacond}, we see that the present NJL prediction is consistent 
with the lattice data,  
confirming that the present model describes the chiral crossover phenomenon quite well.  

\subsubsection{Chiral and axial susceptibility partners}	

\begin{figure}[t]
\centering
\includegraphics[width=0.489\linewidth]{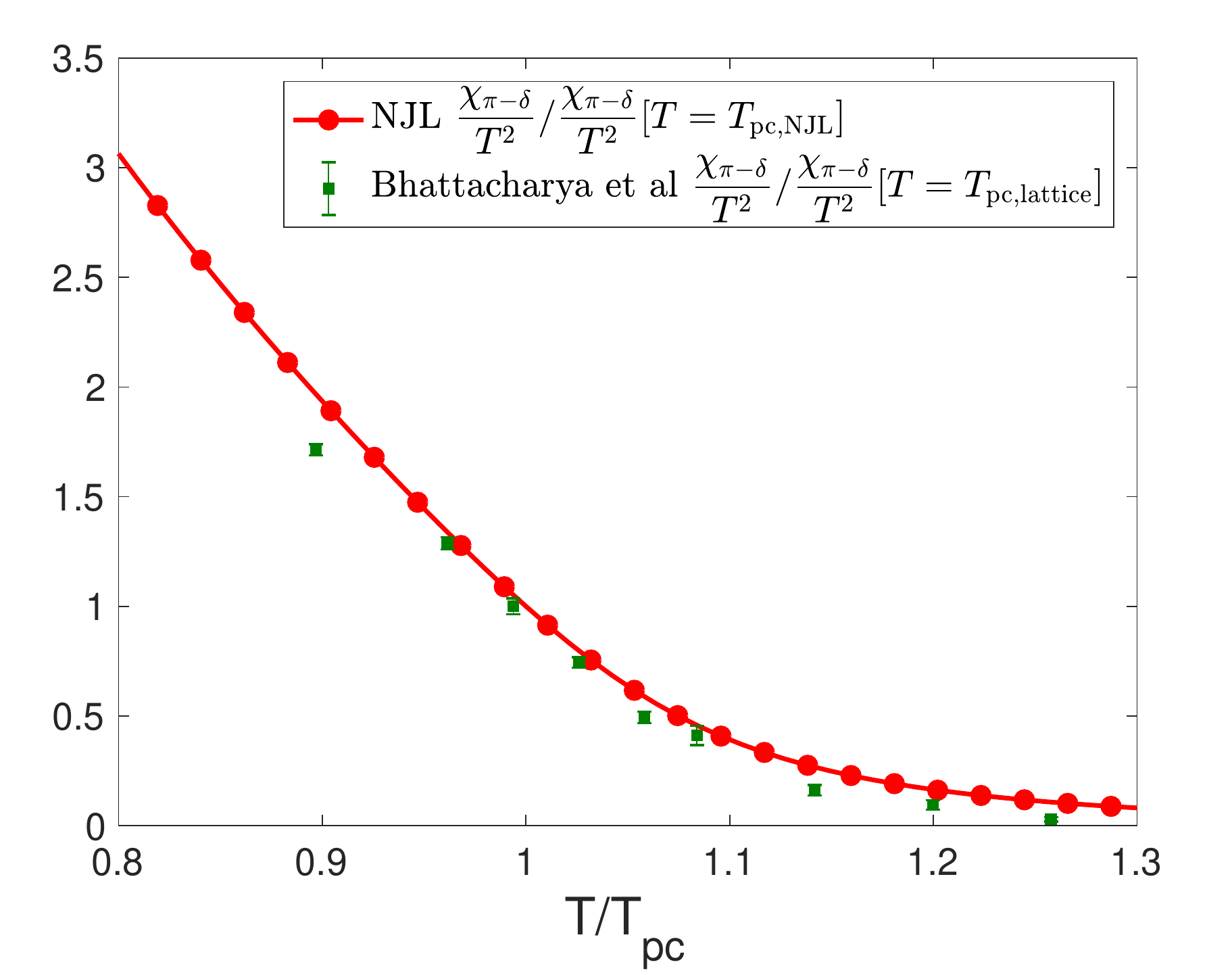}  
\includegraphics[width=0.48\linewidth]{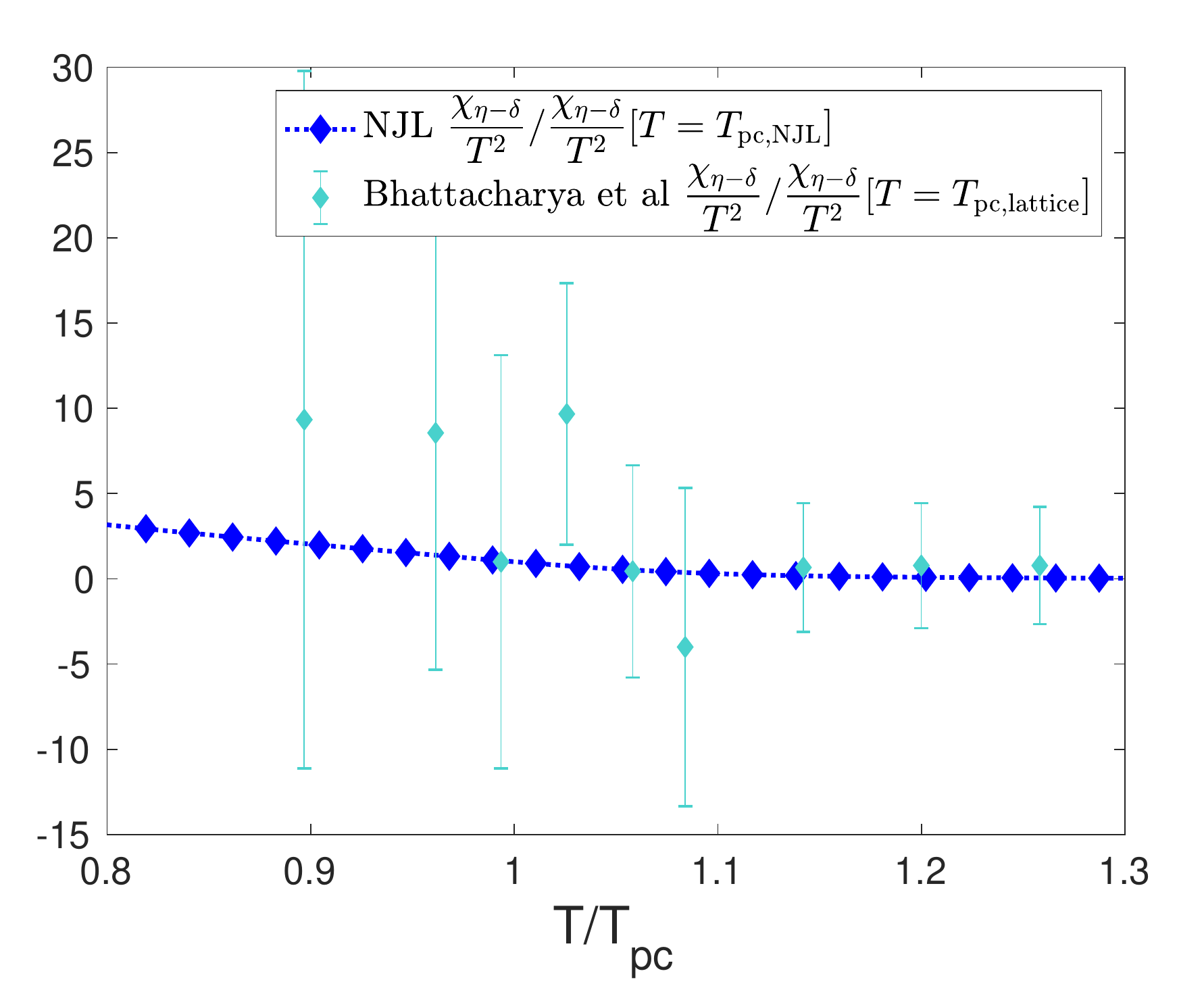}
\caption{The normalized susceptibility differences versus temperature normalized by 
the pseudo-critical temperature $T_{\rm pc}$ for the chiral crossover, in comparison with lattice QCD data for 2 + 1 flavors~\cite{Bhattacharya:2014ara}. The left panel corresponds to $\chi_{\pi- \delta}/T^2$ 
divided by its value evaluated at $T=T_{\rm pc}$, and 
the right panel is the same plots for $\chi_{\eta- \delta}/T^2$. 
The present NJL yields $T_{\rm pc}|_{\rm NJL} = 188$ MeV, and the quoted lattice result predicts 
$T_{\rm pc}|_{\rm lat} = 155$ MeV (at the central value).}
	\label{susdiff}
 \end{figure}

The scalar and pseudoscalar susceptibilities ($\chi_\eta$, $\chi_\pi$, $\chi_\sigma$, $\chi_\delta$) presented 
in Eqs.~(\ref{chip}), (\ref{pion}), (\ref{sigma}), and (\ref{delta}) are correlated with each other by the 
chiral $SU(2)_L \times SU(2)_R$ and $U(1)_A$ transformations~\cite{Bazavov:2012qja}, 
which can be summarized as the following cartoon:  
\begin{center}
\begin{tikzpicture}
  \matrix (m) [matrix of math nodes,row sep=4em,column sep=5em,minimum width=3em]
  {
     \chi_{\pi} & \chi_{\sigma} \\
     \chi_{\delta} & \chi_{\eta} \\};
  \path[-stealth]
    (m-1-1) edge node [midway,left] {$U(1)_A$} (m-2-1)
            edge  node [above] {SU(2)} (m-1-2)
    (m-2-1) edge node [below] {SU(2)} (m-2-2)
    (m-1-2) edge node [right] {$U(1)_A$} (m-2-2)
    (m-2-1) edge node [midway,left] { } (m-1-1)
    (m-1-2) edge node [midway,left] { } (m-1-1)
    (m-2-2) edge node [midway,left] { } (m-1-2)
    (m-2-2) edge node [midway,left] { } (m-2-1);
\end{tikzpicture}
\end{center}
The chiral and axial partners will be degenerate each other in the symmetric limits: 
\begin{align}
    \chi_{\pi} &=\chi_{\sigma},\quad \chi_{\delta}=\chi_{\eta} \qquad \text{(chiral $SU(2)$ symmetric limit)} \notag \\
     \chi_{\pi} &=\chi_{\delta},\quad \chi_{\sigma}=\chi_{\eta} \qquad \text{($U(1)_A$ axial symmetric limit)} 
     \,. 
\end{align} 
Then, observation of null difference between the above partners can effectively monitor the restoration of the related symmetry. 
\textcolor{black}{
Figure \ref{susdiff} shows the plots of $\chi_{\pi-\delta}/T^2$ (left panel) and $\chi_{\eta-\delta}/T^2$ 
(right panel) normalized 
their values measured at $T=T_{\rm pc}$, in comparison to the lattice QCD result~\cite{Bhattacharya:2014ara}. 
Both of the $T/T_{\rm pc}$ dependence on the normalized susceptibilities show qualitative agreement of the present NJL estimates with the current lattice data.  
}

\subsubsection{Topological susceptibility}

\begin{figure}[!htbp]
\centering
    \includegraphics[width=0.8\linewidth]{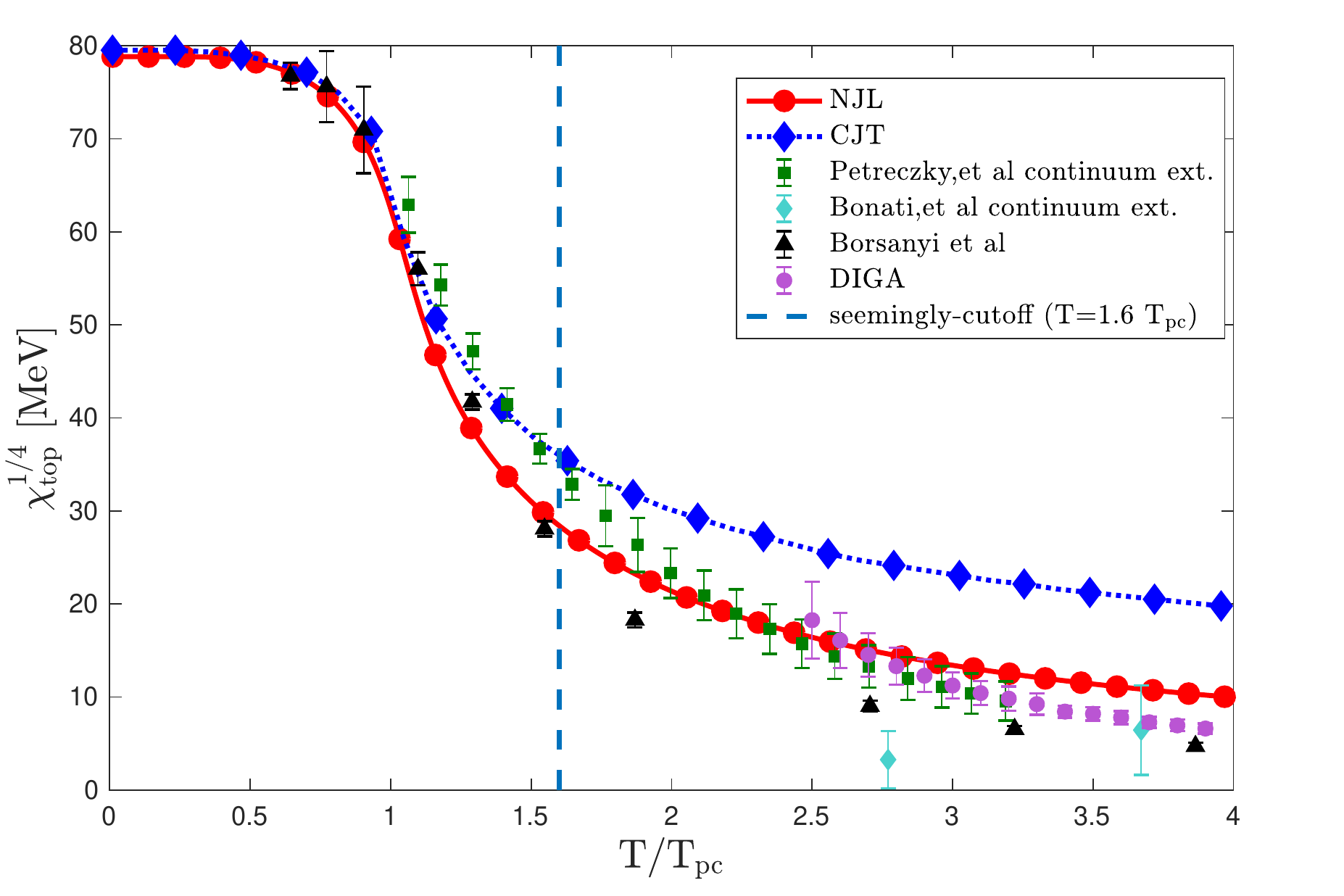}
	\caption{$T/T_{\rm pc}$ dependence of topological susceptibilities,  compared with lattice data~\cite{Petreczky:2016vrs,Bonati:2018blm,Borsanyi:2016ksw} and other models as described in the text.  
	The vertical line at $T/T_{\rm pc} = 1.6$ indicates the theoretical validity of the present NJL model (denoted as the seemingly-cutoff), above which regime the present model description may not be reliable because of lack of the gluonic degrees of freedom (see also footnote \#2). 
	This limit needs to be taken into account in referring to the range of $T/T_{\rm pc}$, when compared to the lattice data displayed in the figure.}  
	\label{chitopfig}
\end{figure}

We numerically evaluate $\chi_{\rm top}$ in Eq.(\ref{chitop}), with the present NJL estimates 
on the quark condensates and pseudoscalar susceptibilities, as a function of temperature. 
In Fig.~\ref{chitopfig}, we plot the temperature dependence of the unnormalized topological susceptibility $\chi_{\rm top}^{1/4}$, where we have 
taken the absolute value of $\chi_{\rm top}$.  
Comparison with the dilute instanton gas approximation (DIGA) \cite{Pisarski:1980md,Gross:1980br}, the linear sigma model result (denoted as CJT in the figure) \cite{Kawaguchi:2020qvg}  and the result from lattice simulation in the continuum limit \cite{Petreczky:2016vrs,Bonati:2018blm,Borsanyi:2016ksw} 
have also been displayed. 
The DIGA prediction has been quoted from the literature~\cite{Petreczky:2016vrs}. 
	For the way of error bars associated with the DIGA, see the cited reference. 
The temperature is normalized by the pseudo-critical temperature in the figure, where 
we have taken $T_{\rm pc}|_{\rm NJL}=188$ MeV for the NJL case, $T_{\rm pc}|_{\rm CJT}=215$ MeV for the linear sigma model case, and $T_{\rm pc}|_{\rm lat}=155 $ MeV for the lattice.

Figure~\ref{chitopfig} shows good qualitative agreement between the NJL analysis and lattice result. We see that even though the NJL description without 
gluonic contribution 
may not be reliable for $T/T_{\rm pc}>$  1.6, the good agreement keeps in the whole range of the available lattice data, $T/T_{\rm pc}\sim0-4$.

At $T<T_{\rm pc}$, all the results fit perfectly with each other, including the linear sigma model estimate. 
In contrast, when $T>T_{\rm pc}$, we see substantial deviation for the linear sigma model prediction from the NJL's and lattice results~\footnote{
\textcolor{black}{
Even within the linear sigma model description, it has been also shown that improved treatment of the thermal characteristics of the linear sigma model based on the application of the functional renormalization group technique to the effective potential reproduces quantitatively correctly lattice results for the pseduocritical temperature~\cite{Fejos:2021yod}. 
}}. 
In the literature~\cite{Kawaguchi:2020qvg}, the pseudoscalar susceptibility terms were not able to evaluate, because the authors did not include the higher order terms in the current quark masses, and therefore, performing the second order derivative on the mass parameter to obtain pseudoscalar susceptibility would not be worked out. 
Thus, their $\chi_{\rm top}$ only includes the quark condensate terms. 
The present NJL model is able to give the pseudoscalar susceptibility contribution to $\chi_{\rm top}$, to achieve an improved estimate on the quark condensate. 
The better qualitative agreement of the NJL with the lattice result may thus imply 
the importance of contributions from 
the pseudoscalar susceptibilities at higher temperatures, though the model estimate may not rigorously be valid beyond the seemingly-cutoff temperature ($T > 1.6 T_{\rm pc}$) as noted above. 

The topological susceptibility $\chi_{\rm top}$ has been discussed 
based on the NJL model descriptions similar to ours~\cite{Fukushima:2001hr,Costa:2008dp,Jiang:2012wm,Jiang:2015xqz}. 
However, the anomalous Ward-identity in Eq.(\ref{WI-def}) 
and the flavor-singlet condition necessary in  
deriving the proper $\chi_{\rm top}$ in Eq.(\ref{chitop}) have been ignored there. 
Absence of the former led to miss-identification of 
the restoration of the axial symmetry, which was played 
by $\chi_{\rm top}$ in the literature, while 
the latter missing factor made improper temperature dependence 
of $\chi_{\rm top}$. 

\section{Evaluation of QCD trilemma estimator}

\subsection{Violation of QCD trilemma at physical point in a whole temperature regime}

\begin{figure}[t] 
  \begin{center}
   \includegraphics[width=10.0cm]{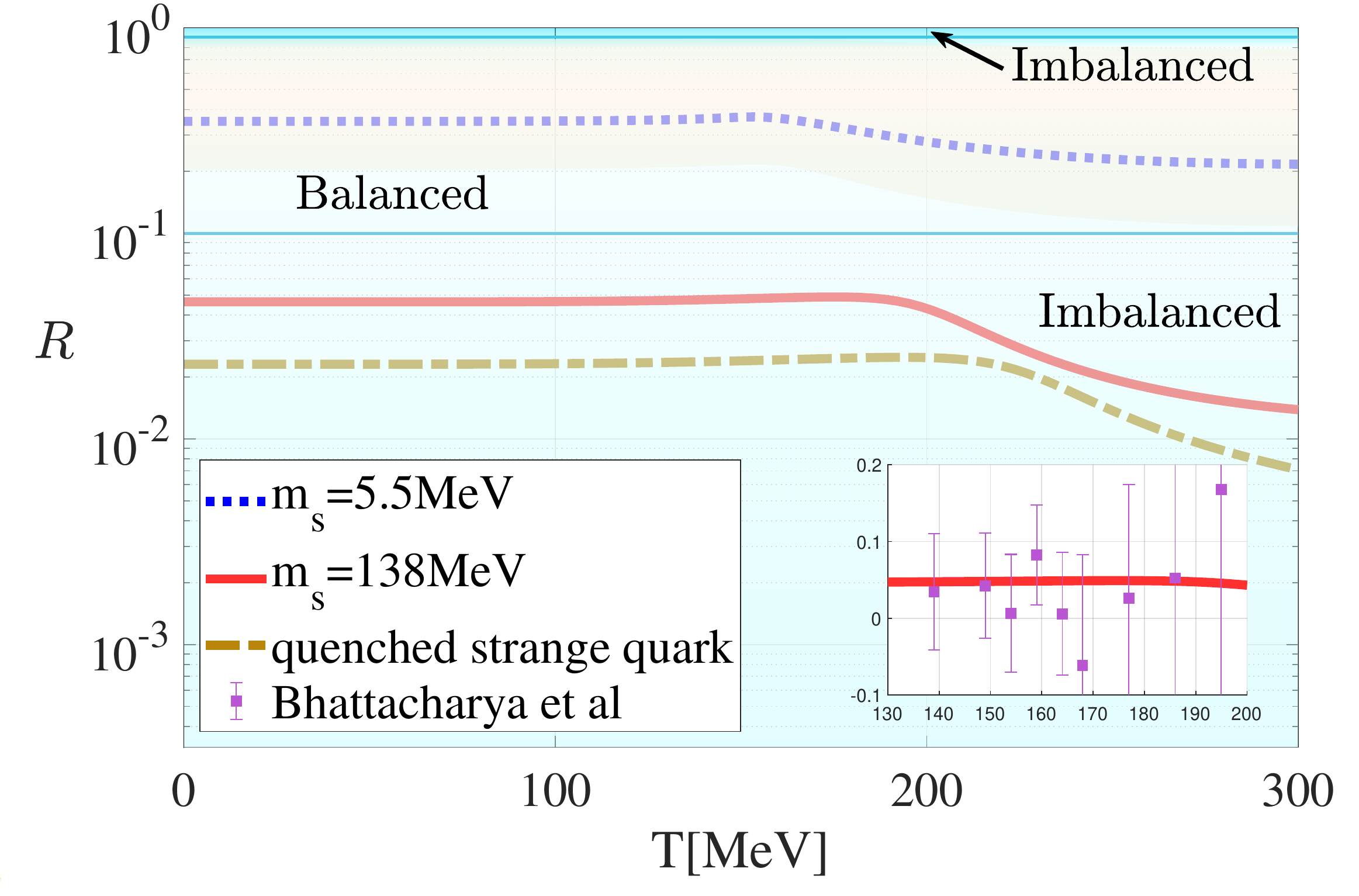}
  \end{center}   
\caption{
Plots showing that real-life QCD is imbalanced, 
which is monitored by the trilemma estimator $R$ defined in Eq.(\ref{R-def}). 
Estimates have been done  
based on the NJL model described as in the text. 
Comparison with the $2 + 1$ flavor-lattice QCD data (with $m_\pi=135$ MeV) in the available $T$ range 
has also been displayed with the error bars~\cite{Bhattacharya:2014ara} (in the zoomed-in window). 
The curve with $m_s=138$ MeV points to real-life QCD with 
three flavors, while the quenched-strange quark limit has been achieved by taking   
$m_s=50$ GeV, corresponding to   
the two-flavor limit. Another curve with $m_s=5.5$ MeV denotes 
a conjectured prediction in the three-flavor symmetric limit. 
The ``balanced" and ``imbalanced" regimes are defined in Eq.(\ref{natural-region}). 
}  
\label{Ratio-Tdep}
\end{figure}

Figure~\ref{Ratio-Tdep} shows values of 
the trilemma estimator $R$ evolved with $T$, 
allowing $m_s$ off the physical point with $m_l$ kept physical. 
See the middle-solid curve with $m_s=138$ MeV, which corresponds 
to real-life QCD.  
Comparison with the available $2+1$ flavor-lattice QCD data (with 
$m_\pi = 135$ MeV) on $R$~\cite{Bhattacharya:2014ara}  --- reconstructed from the data on 
$\chi_{\pi -\delta}$ and $\chi_{\eta - \delta}$ through the relation Eq.(\ref{WI-R}) --- has also been displayed (in the zoomed-in window), 
which shows good agreement including the error bars, for 
$140 \,{\rm MeV} \lesssim T \lesssim 200\,{\rm MeV}$. \textcolor{black}{The reconstructed data of $R$ include large errors, which is mainly due to the large uncertainty of the lattice measurement on $\chi_{\eta-\delta}$ (See Fig.~\ref{susdiff}).}

Remarkably, in a whole temperature regime including the chiral crossover regime, 
real-life QCD stays outside the ``balanced" region defined as in Eq.(\ref{natural-region}). 
We have observed $R \simeq 0.05$ at around $T$ covering the  
crossover point ($T_{\rm pc}|_{\rm NJL} \simeq 188$ MeV: 
$140 \,{\rm MeV} \lesssim T \lesssim 200\,{\rm MeV}$), consistently  with the lattice data, 
and  
$R \lesssim 0.01$ at $T \gtrsim 300$ MeV. 
Namely, the amount of imbalance is slightly amplified by thermal loop effects as $T$ develops from zero~\footnote{
Above $T \sim 300$ MeV corresponding to the typical scale of the constituent quark mass, the NJL description as the effective theory of QCD 
may be somewhat unreliable because the deconfining color degrees of freedom and 
thermal gluonic contributions would be significant. }.

One might note that subtraction by $\chi_\delta $ in Eq.(\ref{WI-def}) 
is ambiguous, and can be replaced by another chiral 
susceptibility in the sigma meson channel ($\chi_\sigma$). 
We have checked that 
this replacement does not alter our main conclusion that 
real-life QCD involves big imbalance. 
We have also found that $\chi_{\pi-\sigma} \gg \chi_{\eta - \sigma}$ at $T=0$, 
$\chi_{\pi-\sigma} \ll \chi_{\eta - \sigma}$ at around the chiral crossover, 
then $\chi_{\pi-\sigma} $ will get close to $\chi_{\pi -\delta}$, and finally go to zero. 
The latter trend is consistent with the currently available 
lattice data~\cite{Bhattacharya:2014ara}.

Although the present model parameters are fixed at 
the physical point, we may deduce some conjectures 
on the violation of QCD trilemma in a view of the quark mass difference. 
Extrapolating off real-life QCD,   
one can then observe that the ``imbalanced" domain still covers 
the two-flavor limit case with $m_s = 50$ GeV (bottom-dot-dashed curve), where strange 
quark is decoupled, and the amount of imbalance is greater than that in 
the real-life QCD case. 
Taking the three-flavor symmetric limit $m_s=m_l$ with $m_l$ 
fixed to the physical value, 
we find ``balanced" QCD (top-dashed curve), 
which keeps almost constant $R$ at any finite $T$ 
within the ``balanced" interval in Eq.(\ref{natural-region}). 
This implies that the three-flavor symmetry would be related to 
the relaxation of the QCD trilemma.

\subsection{QCD trilemma and flavor symmetry}

\begin{figure}[t]
  \begin{center}
   \includegraphics[width=9.5cm]{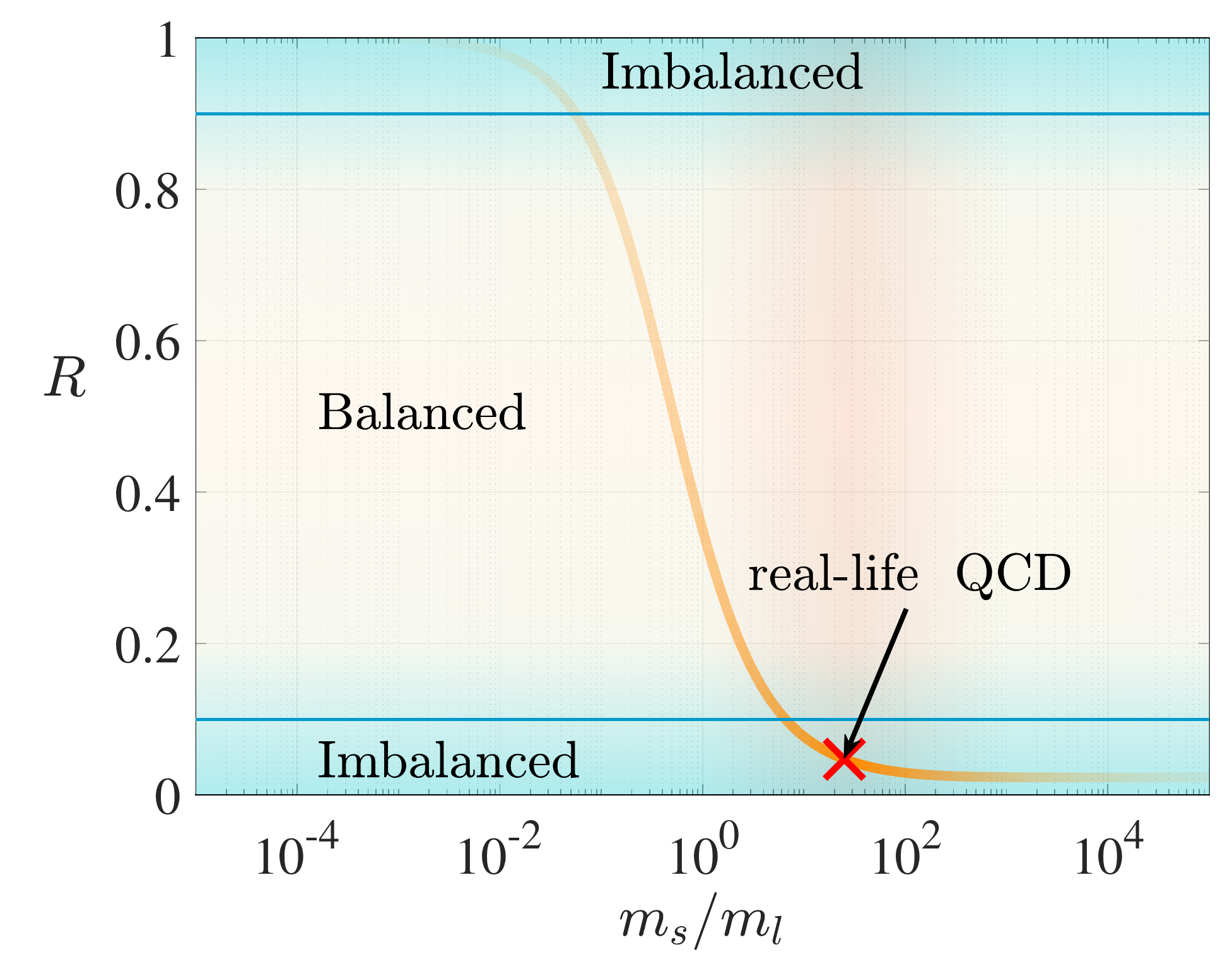}
  \end{center}   
\caption{
Plots on the QCD trilemma estimator $R$ at $T=0$ as a function of $m_s/m_l$, 
along with the ``balanced" interval defined in Eq.(\ref{natural-region}). 
The shaded domain surrounded by the real-life QCD point 
implies confidence level intervals for the model 
prediction, where the model parameters, except for 
$m_s$, have been fixed at the physical point, as 
noted in the text. 
The thinner-shaded regions should be understood as 
indefinitely extrapolated results with somewhat poor 
reliability. 
}  
\label{Ratio-mdep}
\end{figure}

Since the order of magnitude for $R$ tends to be almost fixed at $T=0$, 
we may focus only on $R$ at $T=0$, and look into the 
flavor-symmetry dependence on $R$, 
by varying $m_s$ in a wide range, with fixed $m_l$ to the 
physical value. 
Figure~\ref{Ratio-mdep} shows plots on $R$ as a function of $m_s/m_l$, 
together with the ``balanced" interval in Eq.(\ref{natural-region}).  
As $m_s$ goes off the flavor symmetric limit in the ``balanced" domain 
to be smaller, 
$R$ tends to get larger, to flow into the ``imbalanced" domain 
with gigantically suppressed $\chi_{\rm top}$.
The figure clearly shows that 
``balanced" QCD should have had some approximate  
three-flavor symmetry for up, down and strange quarks with 
$0.06 \lesssim m_s/m_l \lesssim 6$. 

\begin{figure}[t] 
\centering
\includegraphics[width=0.47\linewidth]{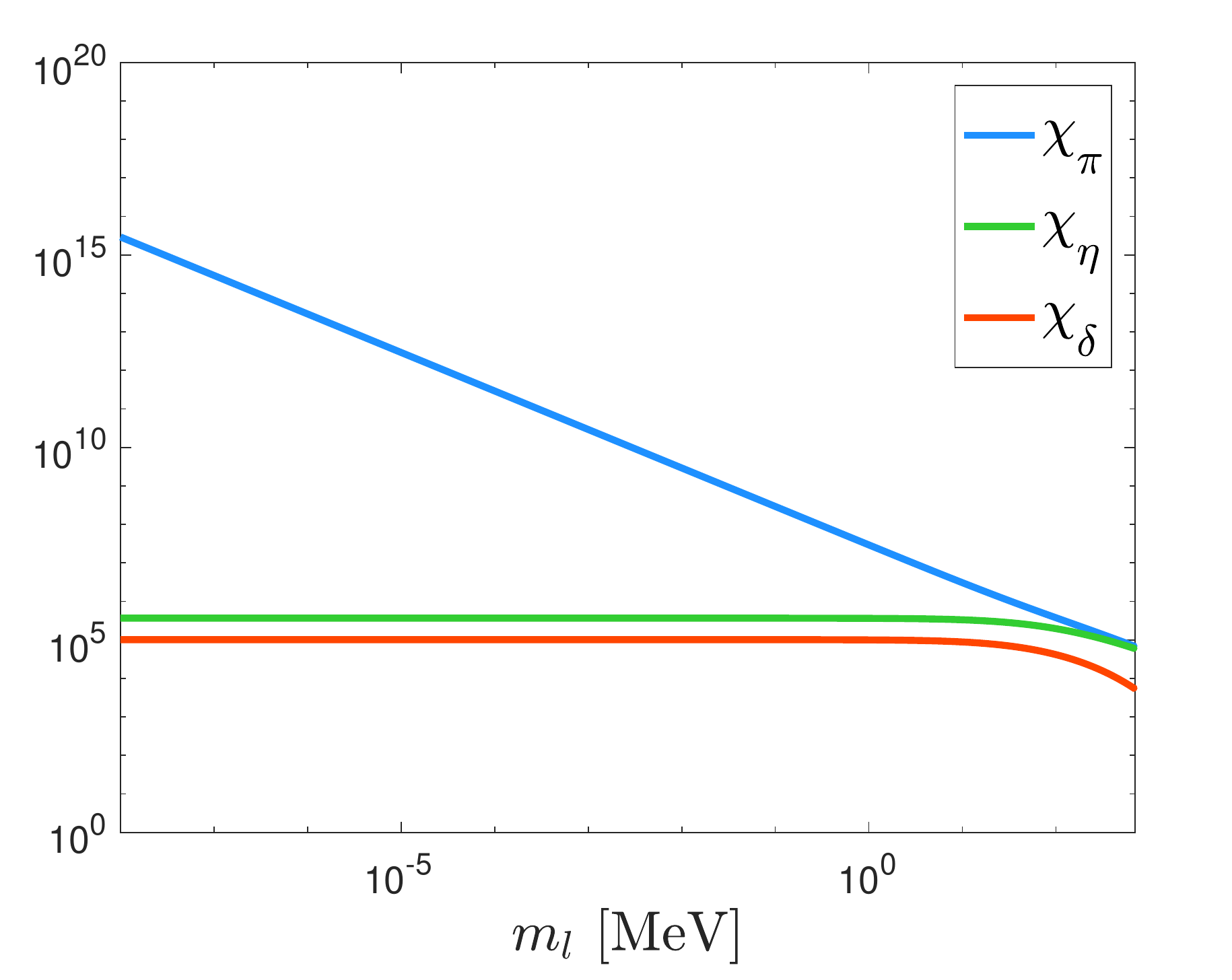}  
\includegraphics[width=0.5\linewidth]{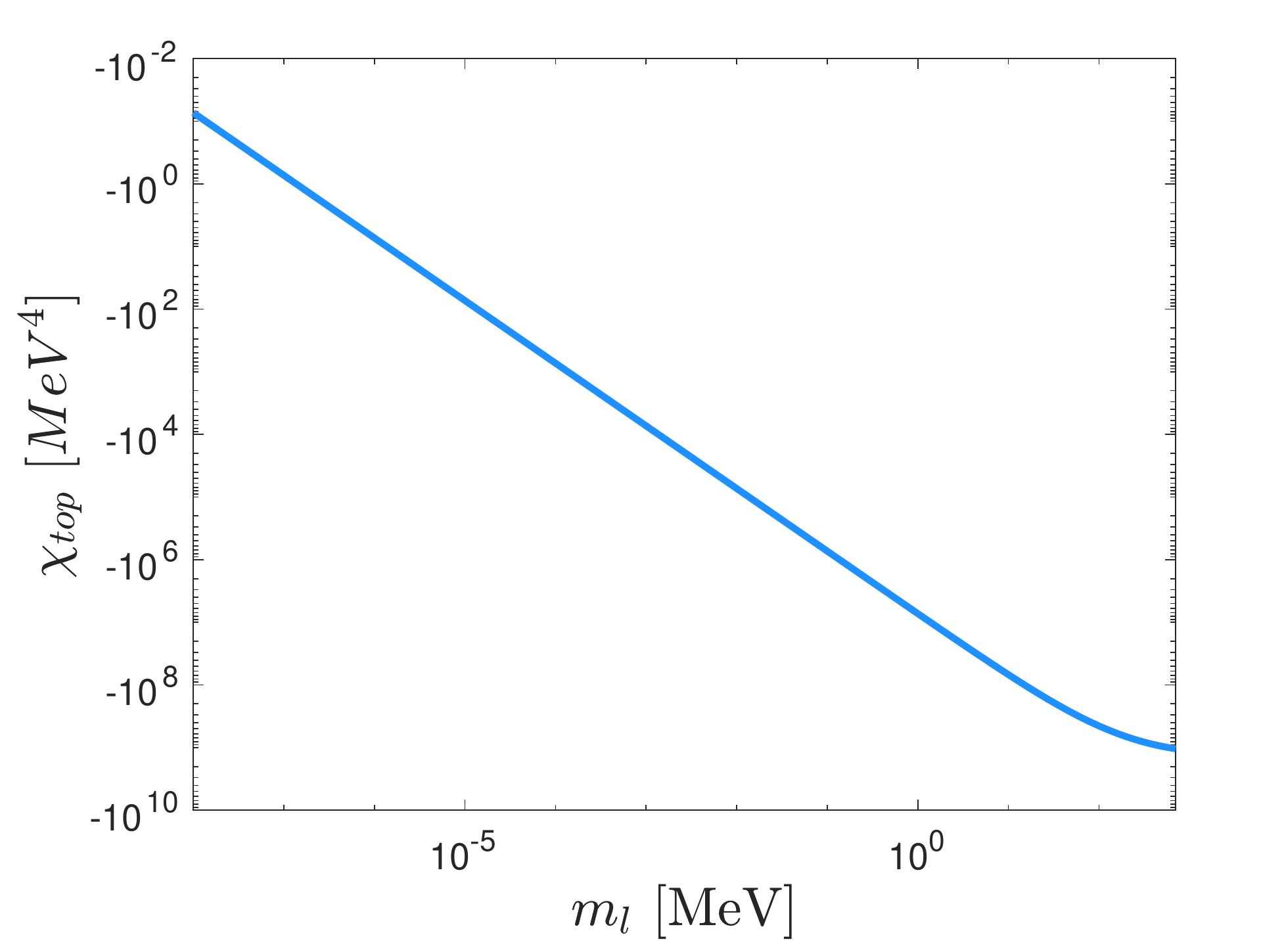}
\caption{
(Without the three-flavor symmetry): the $m_l$ dependence on $\chi_\pi, \chi_\eta$, $\chi_\delta$ [MeV$^2$] (left panel), and 
$\chi_{\rm top}$ [MeV$^4$] (right panel) 
at $T=0$. The value of strange quark mass $m_s$ is fixed to be equal to $m_l$ in the plots.}
\label{susc_ml}
\end{figure}

We shall investigate the dependence of the flavor-symmetry violation 
on the imbalanced QCD trillemma in more details. 
\textcolor{black}{
First of all, we may simply suppose that 
the scalar and pseudoscalar susceptibilities 
are scaled with the associated meson masses~\footnote{
The susceptibilities correspond to meson-correlation functions 
at zero momentum transfer. 
This is in contrast to the conventional meson correlators depending on the transfer momentum, 
from which meson masses are read off.
Furthermore, 
the susceptibilities involve contact term contributions independent of momenta, which could be sensitive to a high-energy scale physics, while the conventional meson correlators are dominated by the low-lying meson mass scale. 
Nevertheless, the degeneracy of the chiral or axial partners at high temperatures, similar to those detected in the susceptibility, can also be seen in the mass difference or equivalently the degeneracy of the conventional meson correlators for the partners, which is simply because the mass difference plays an alternative indicator of the chiral  or axial breaking 
\textcolor{black}{as observed in the lattice simulations~\cite{Brandt:2016daq,Brandt:2019ksy}.}  
}, like 
$\chi_\delta \propto 1/m_{\delta}^2$, $\chi_\eta \propto 1/m_{\eta}^2$ and 
$\chi_\pi \propto 1/m_{\pi}^2$, 
and consider the light quark mass $m_l$ to generically differ from the strange 
quark mass $m_s$, including the real-life QCD case with the three-flavor symmetry broken. 
Among the susceptibilities, 
$\chi_\pi$ is most sensitive to the current mass of the light quarks ($m_l$), 
because the pion is the pseudo Nambu-Goldstone boson of spontaneous breaking of $SU(2)_L \times SU(2)_R$ symmetry 
carried by the light quarks. The $\chi_\pi$ thus monotonically gets smaller (larger), 
as $m_l$ gets larger (smaller), by following $\chi_\pi \propto 1/m_{\pi}^2 \sim 1/m_l$. 
On the other hand, the other pseudoscalar susceptibility $\chi_\eta$ significantly involves the U(1)-axial anomaly contribution in $m_\eta$,}  
\textcolor{black}{so it almost keeps constant in $m_l$.}
\textcolor{black}{The scalar susceptibility $\chi_\delta$, free from the Nambu-Goldstone boson nature,} 
\textcolor{black}{also keeps constant with the change of $m_l$.}
\textcolor{black}{ 
Besides, the topological susceptibility $\chi_{\rm top}$ also simply scales with $m_l$, respecting 
the flavor-singlet condition: $\chi_{\rm top} \to 0$ as $m_l \to 0$, 
and will be completely constant in $m_l$ for $m_l > m_s$ due to decoupling of the ``light" quarks. 
Thus the difference in magnitude of susceptibilities are simply originated from the scaling properties with respect to the current mass of 
the light quarks. }
\textcolor{black}{We plot those $m_l$ scaling behaviors (at $T=0$) in Fig.~\ref{susc_ml}.} \textcolor{black}{The light quark mass is allowed to vary from $10^{-2}$eV to the cutoff scale of the presently employed NJL model (631.4 MeV), since the result from $m_l$ above the cutoff scale would be of poor reliability.} 
\textcolor{black}{From the figure, the $m_l$ dependence 
is read off and the susceptibilities are 
found to take simple power laws when $m_l \lesssim m_s$: }
\textcolor{black}{
\begin{align} 
 & \chi_\pi \sim m_l^{-1} 
\,, \notag\\ 
 & \chi_\eta \sim 
 {\rm constant}\,,  \quad {\rm for} \qquad m_l < m_s \,, \notag\\
 & \chi_\delta \sim 
 {\rm constant}\,,  \quad {\rm for} \qquad m_l < m_s  
\,, \notag\\ 
 & \chi_{\rm top} \sim m_l
  \quad {\rm for}\,, \qquad m_l < m_s
\,. \label{power:sus}
\end{align}  
}

\begin{figure}[t] 
\centering
\includegraphics[width=0.483\linewidth]{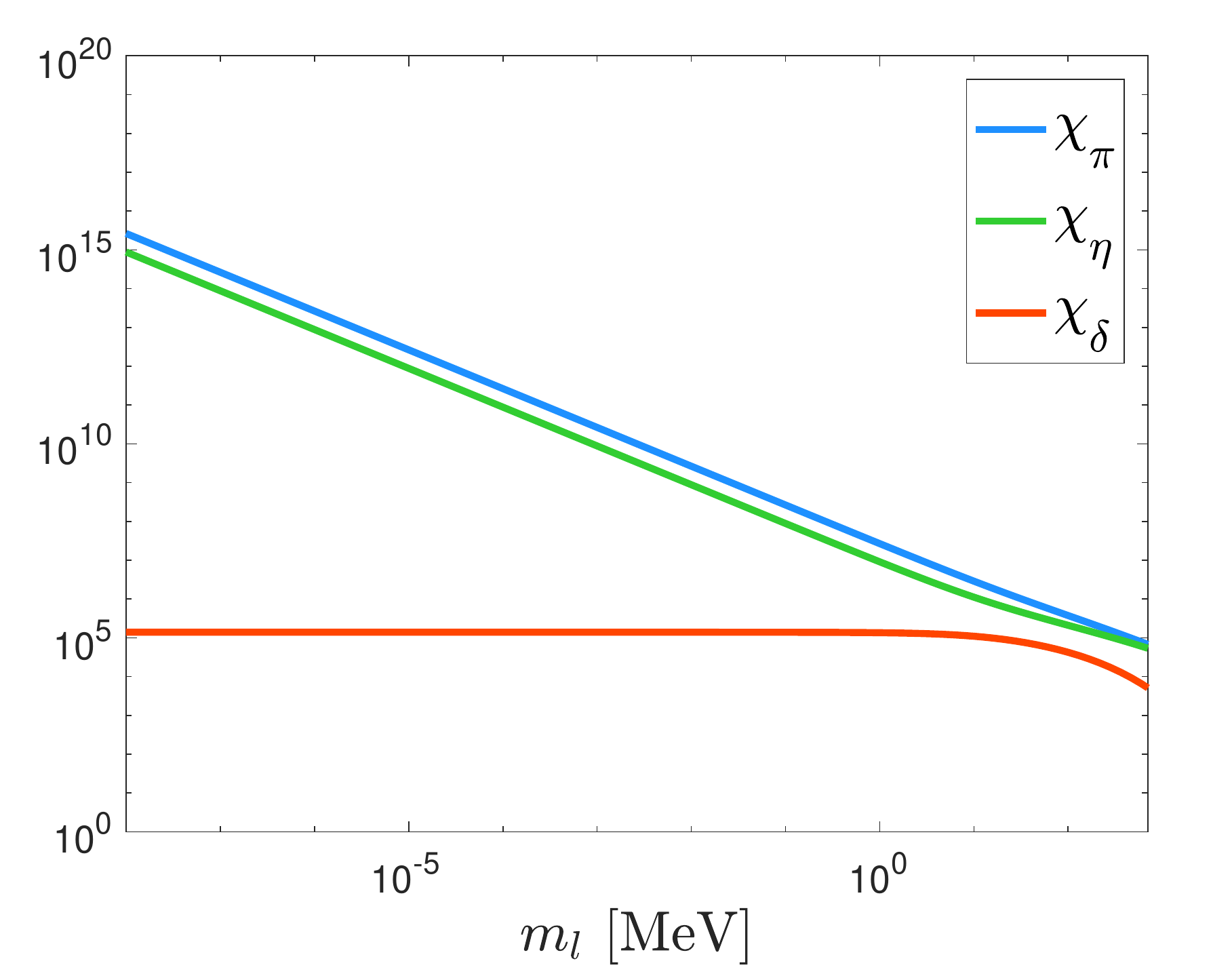}  
\includegraphics[width=0.5\linewidth]{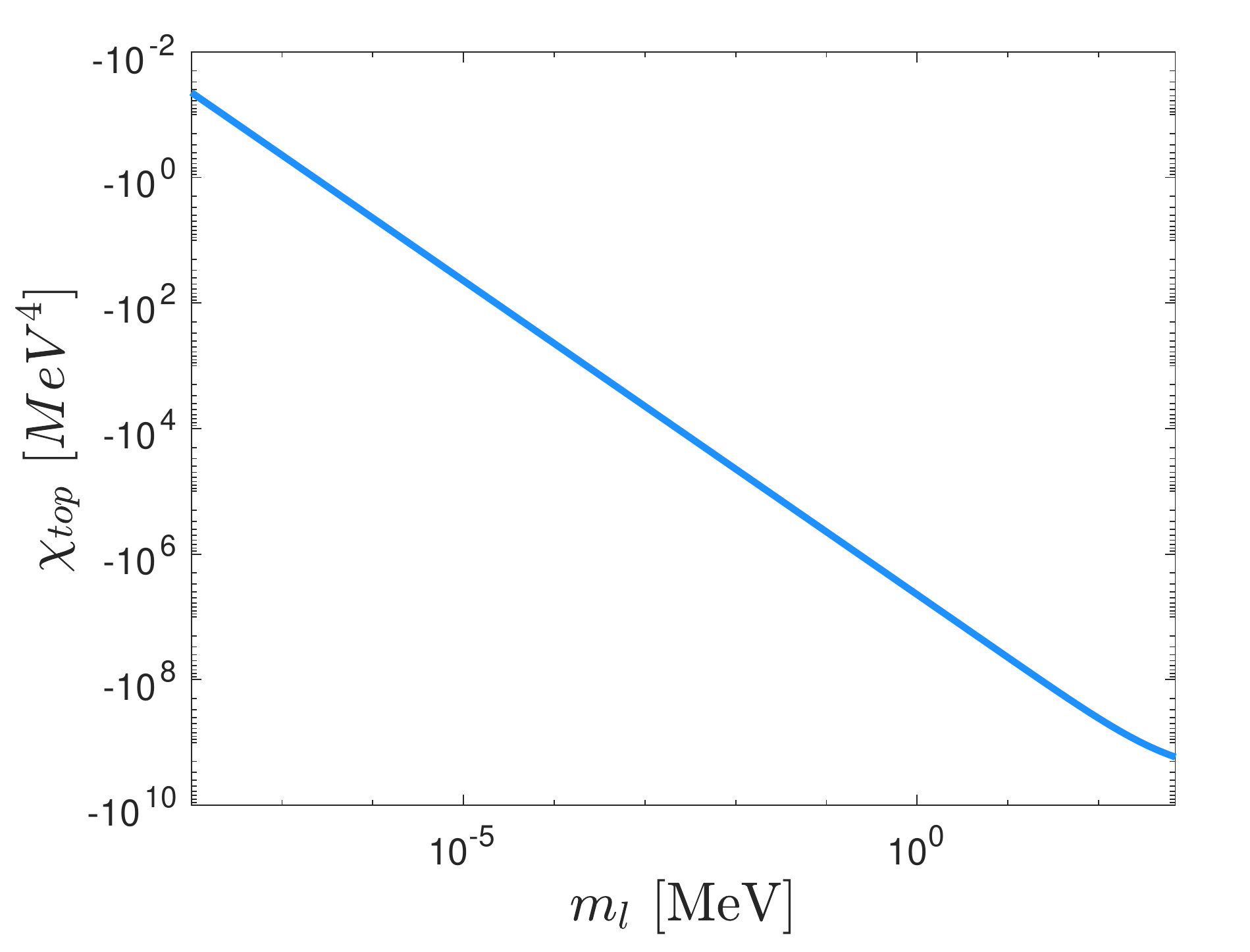}
\caption{
(With the three-flavor symmetry): 
The same plots as those in Fig.~\ref{susc_ml}, but with the three-flavor symmetry, where  
$m_s = m_l$. }
\label{susc_ml_ms}
\end{figure}

 Next, consider the three-flavor symmetric limit, where $m_l$= $m_s$. 
In this case QCD is balanced as noted above.  
It also turns out that the scaling law of $\chi_\eta$ in Eq.~(\ref{power:sus}) is broken: 
the Ward identity Eq.~(\ref{disc}) 
tells us that  
the difference between $\chi_\pi$ and $\chi_\eta$ is controlled by the $4\frac{m_s}{m_l}\chi_P^{ls}$ term 
(where $\chi_P^{ls} = \chi_P^{us} = \chi_P^{ds}$).
Since no preference among quark flavors is present in the flavor symmetric case,  
$\chi_P^{ls}$ should be on the same order of magnitude as that of $\chi_\pi$, 
which we have indeed numerically confirmed. See Fig,~\ref{susc_ml_ms}. 
Since $m_l=m_s$, there is no extra power scaling of $1/m_l$ 
which is present in the flavor asymmetric case and leads to big enhancement of 
the ($4\frac{m_s}{m_l}\chi_P^{ls}$) part to destructively interfere with $\chi_\pi$, 
yielding a much suppressed $\chi_\eta$ compared to $\chi_\pi$  
(See Fig.~\ref{susc_ml}). 
Thus the scaling law of $\chi_\eta$ is the same as that of $\chi_\pi$, i.e., $\chi_\eta$ $\sim$ $m_l^{-1}$, while others take the same scaling laws as in Eq.(\ref{power:sus}), namely, \begin{align} 
 & \chi_\pi \sim \chi_\eta  \sim m_l^{-1} 
%
\quad {\rm for} \qquad m_l = m_s 
\,, \notag\\
 & \chi_\delta \sim 
 {\rm constant}  \quad {\rm for} \qquad m_l = m_s  
\,, \notag\\ 
 & \chi_{\rm top} \sim m_l
  \quad {\rm for} \qquad m_l = m_s
\,, \label{power:sus:sym}
\end{align}  
as depicted in Fig.\ref{susc_ml_ms}.

\textcolor{black}{This scaling violation in the flavor symmetric case 
can also be understood as a big suppression of the $U(1)_A$ 
anomaly contribution, coupled to the flavor violation, to $m_\eta^2$, which dominates 
in $\chi_\eta$ in the flavor asymmetric case: 
in the flavor symmetric case we have 
$\chi_\pi = \chi_P^{88}$, and $\chi_\eta = \chi_\pi + 4 \chi_P^{ls}$ 
with $\chi_P^{ls} =1/6 (\chi_P^{00} - \chi_P^{88})$. 
Straightforward numerical evaluation reveals that 
$\chi_P^{88} \gg \chi_P^{00}$ for small $m_l$. 
Then, we find 
$\chi_\eta \approx \chi_\pi/3 \sim 1/m_l$ for small $m_l$.  
In particular, note that $\chi_P^{88} = \chi_\pi$ does not include 
the $U(1)_A$ anomaly effect, and is now much larger than 
the $U(1)_A$-anomaly affected $\chi_P^{00}$ part, 
which implies the $U(1)_A$ anomaly contribution is much suppressed in 
$\chi_\eta$, hence in $m_\eta$ as well. 
}

Thus the three-flavor symmetry brings the $m_l$ scaling universal 
among susceptibilities for the pion and eta meson, by 
suppressing the $U(1)_A$ anomaly contribution only in the eta meson 
channel. 
This, on the other side of the same coin, indicates 
that the amplification of the $U(1)_A$ anomaly effect in $\chi_{\eta - \delta}$ 
so that the chiral symmetry is made faster restored than the axial symmetry 
with holding the anomalous chiral Ward identity Eq.(\ref{WI-def}).




\section{Conclusion}

In conclusion, 
real-life QCD is required to relax the trilemma $(R \ll 1)$, 
meaning that the much smaller strength of the chiral breaking is given  
by a big cancellation between the strength of the axial breaking  and 
the transition rate of the topological charge. 
This is schematically 
depicted in Fig.~\ref{QCD-trilemma}. 
This is ``imbalance" of the QCD vacuum, present 
in a whole temperature regime of thermal QCD including the vacuum at $T=0$. 
This imbalance or violation of QCD trilemma is triggered 
due to the 
three-flavor symmetry violation for up, down and strange quarks, 
in particular the lightness of up and down quarks. 
The violation of the three-flavor symmetry 
specifically brings enhancement of the $U(1)_A$ anomaly contribution in the chiral $SU(2)$ breaking to be insensitive to the smallness of the light quark mass, 
while the axial indicator and the transition rate of the topological 
charge are fairly insensitive to the flavor 
symmetry --- in other words, the flavor symmetry 
dramatically suppresses the $U(1)_A$ anomaly effect in the chiral $SU(2)$ breaking. 
This implies that in the QCD vacuum with the trilemma realized,  
all the three-flavor octet pseudoscalar mesons act like identical pseudo Nambu-Goldstone bosons, 
so that the chiral 
and axial breaking together with the topological 
charge contribute to the QCD vacuum on the same order of 
magnitude.

The present work 
confirms and extends the suggestion recently reported from lattice QCD with 
2 flavors on  
dominance of the axial and topological susceptibilities left in the chiral susceptibility at 
high temperatures over the chiral crossover~\cite{Aoki:2021qws}.

The violation of QCD trilemma 
would be crucial to deeply pursue 
the expected dominance of the chiral symmetry 
breaking in the origin of mass, and provides the understanding of 
mechanism how the faster (effective) restoration of the chiral symmetry 
in the presence of contamination with the $U(1)_A$ anomaly is achieved: 
it is understood as a big cancellation between the axial and topological 
susceptibilities, due to the three-flavor violation among 
up, down and strange quarks. 
Our findings can directly be tested on lattice QCD with 2 + 1 flavors 
at physical point and also off physical point, in the future.

It would be nice if the violation of QCD trilemma could be evaluated directly using lattice QCD, but since the quark masses need to be varied, the lattice calculation is so costly. As the first step, it would be appropriate to perform the calculation using a conventional effective model like the present NJL. 
In fact, it has been shown that the present NJL model matches 
the lattice results on a couple of observables 
relevant to the chiral crossover regime, 
within a deviation of about 30\% (consistent with the expected 
theoretical uncertainty of the large $N_c$/mean field approximation on that the present NJL is built). 
This shows that the present NJL model, though being based on 
the rough mean field approximation, 
indeed is even quantitatively valid as an effective model of QCD, as good as the lattice QCD. 
Thus the present work gives motivation to the research in 
other QCD-compatible theories, such as lattice and/or functional renormalization group approach.

In closing, we give several comments related to possible applications of 
concept
of QCD trilemma:

\begin{itemize} 

\item 
The notion of QCD trilemma and its violation would also provide 
us with a new guiding principle to explore the 
flavor dependence of the chiral phase transition, 
such as the Columbia plot~\cite{Brown:1990ev}.

\item 
It would be also intriguing to study the violation of QCD trilemma, 
by means of the nonperturbative renormalization group.

\item 
Correlations between the violation of QCD trilemma 
and deconfinement-confinement phase transition 
can be addressed, when the present NJL model is extended 
by including the Polyakov loop terms.

\item 
Since the form of the anomalous chiral Ward identity in Eq.(\ref{WI-def}) 
will be intact as long as the quark mass terms are 
only the leading source to explicitly break the three-flavor chiral 
symmetry, the presently described argument would possibly be applicable also to 
dense QCD, and thermomagnetic QCD, where in the latter case the electromagnetic interactions contribute  
as the subleading (loop) order to the chiral breaking, as in Eq.(\ref{rot-O}).

\end{itemize} 

Those will deserve to another publication.

\section*{Acknowledgement}
We thank Tetsuo Hatsuda for giving us a crucial comment on the 
susceptibilities,  
and are also grateful to Hidenori Fukaya for fruitful discussion. 
This work was supported in part by the National Science Foundation of China (NSFC) under Grant No.11747308, 11975108, 12047569, 
and the Seeds Funding of Jilin University (S.M.).  
The work of A.T. was supported by the RIKEN Special Postdoctoral Researcher program
and partially by JSPS  KAKENHI Grant Number JP20K14479.

\appendix 

\section{NJL formulae}
\label{App:formulas}

In this Appendix we list the NJL formulas used for the inputs observables in Eq.(\ref{inputs}), 
which can be found in Ref.~\cite{Hatsuda:1994pi}: 
	\begin{itemize}
	\item The pion decay constant $f_{\pi}$: it is computed by directly evaluating quark loop contributions to the spontaneously broken $SU(2)$ axial current $J_\mu^{A, a}= \bar{l} \gamma_\mu \gamma_5 (\sigma^a/2) l$ ($a=1,2,3$),  
	through the definition of $f_\pi$, $\langle 0| J_\mu^{A \, a}(0) | \pi^b(p) \rangle  = - i p_\mu f_\pi \delta^{ab}$. 
	In the NJL model with 
	the large $N_c$ limit taken (summing up the ring diagrams), we thus have 
		
		\begin{equation}
			f_{\pi} =G_{\pi q}(m_{\pi})M_{u}\frac{2 N_c}{\pi^2}\int^{\Lambda}_{0} dp \frac{p^2}{\sqrt{M_u^2 + p^2}[4(M_u^2+p^2)-m_{\pi}^2]}
			\,, 
			\end{equation}
		where $G_{\pi q}(p=m_{\pi})$ is the pion wavefunction renormalization amplitude evaluated at the onshell, 
		\begin{equation}
			G^2_{\pi q}(m_{\pi})=\left (\frac{N_c}{2\pi^2} \int^{\Lambda}_0 p^2 dp \frac{\sqrt{M_u^2+p^2}}{(p^2+M_u^2-\frac{m_{\pi}^2}{4})^2} \right )^{-1}
		\,. 
		\end{equation}

		\item The pion mass $m_\pi$ is computed by extracting the pole of the pion propagator dynamically generated by the quark loop contribution in the NJL with a resummation technique (Random Phase Approximation) applied~\cite{Hatsuda:1994pi}. The pole position is thus detected as 
		\begin{equation}
			1+G_{\pi}\Pi_{\pi}(m_\pi^2)=0
		\,.  
		\end{equation}
		This pion mass is actually related to the light quark condensate, via the 
		the low energy theorem (the so-called Gell-Mann-Oakes-Renner relation): 
		\begin{equation}
			m_{\pi}^2 =-2m_l\langle \bar ll  \rangle/f_{\pi}^2
			\,.  
			\end{equation}
		
	\item The kaon mass $m_{K}$ is calculable in the same way as in the case of $m_\pi$ above: 
	
		\begin{equation}
			1+G_{K}\Pi_{K}(m_K^2)=0
		\,, 
		\end{equation}
		where
		\begin{equation}
			G_K=g_s + g_D \langle \bar u u \rangle
		\,, 
		\end{equation}
		\begin{equation}
			\Pi_K(w)=2F(w; u,s)+2F(-w; s, u)
		\,, 
		\end{equation}
		\begin{equation}
			F(w; i,j)=-\frac{N_c}{4\pi^2}\int^{\Lambda}_0 p^2 dp \left [ \frac{1}{E_{i}}f_{ij}(w) + \frac{1}{E_{j}}f_{ji}(w) \right ] 
		\,, 
		\end{equation}
		\begin{equation}
		f_{ij}(w)=2\frac{M_i(M_j-M_i)-E_iw}{E_j^2-(E_j+w)^2}
		\,. 
		\end{equation}

	\item The $\eta'$ mass $m_{\eta^{\prime}}$ is identified as the highest mass eigenvalue arising 
	from the mass mixing in the $0-8$ channel. 
	Similarly to the pion and kaon cases, 
	$m_{\eta'}$ is then extracted by the highest pole of the mixed propagator in the $0-8$ channel, $D(q^2),$ as 
		
		\begin{equation}
		D_{\eta^{\prime}}(m_{\eta^{\prime}}^2)=0
		\,, \end{equation}
		where
		\begin{equation}
			D(q^2)=-G_P^{-1}\left ( \frac{1}{1+G_P \Pi^P(q^2)} \right ) \equiv \begin{pmatrix}
						A(q^2) & B(q^2)\\
						B(q^2) & A(q^2)
				\end{pmatrix}		
\,, 
		\end{equation}
through the diagonalization process like 
\begin{equation}
	T_{\theta}D^{-1}(q^2)T_{\theta}^{-1} =
	\begin{pmatrix}
		D_{\eta^{\prime}}^{-1}(q^2) & 0 \\
		0 & D_{\eta}^{-1}(q^2)
	\,\end{pmatrix}
	\,. 
\end{equation}
Here $\Pi^P(q^2)$ is a function given as the generalization of Eq.~(\ref{Pi-p}) with the replacement of the loop function $I_{ii}^P(w^2)$:
\begin{equation}
	I_{ii}^P(w^2)=-\frac{4N_c}{\pi^2}\int^{\Lambda}_0 p^2 dp \frac{E_i}{4E_i^2-w^2}
\,,  
\end{equation}
and $T_{\theta}$ is the diagonalization matrix, 
\begin{equation}
	T_{\theta}=
	\begin{pmatrix}
		\cos(\theta) & -\sin(\theta) \\
		\sin(\theta) & \cos(\theta)
	\end{pmatrix}
	, \quad \tan(2\theta)=\frac{2B(q^2)}{C(q^2)- A(q^2)}
\,. 
\end{equation}
\end{itemize}

\end{document}